\documentclass{amsart}
\usepackage{geometry}
\usepackage{amsmath,amsthm,amssymb,amsxtra,latexsym,mathrsfs,url} % 

\usepackage{ab} %

\usepackage{cite} % bibtex 

\usepackage{tikz-cd} % for commutative diagrams using tikz

\theoremstyle{plain}
\newtheorem{proposition}{Proposition}[section] % reset the counter in each new section (in rest.tex: for each of the four parts)
\newtheorem{lemma}[proposition]{Lemma} % "[proposition]" tells latex to number lemmas and propositions using the same counter
\newtheorem{theorem}[proposition]{Theorem} 
\newtheorem{corollary}[proposition]{Corollary}
\newtheorem{method}[proposition]{Method}
\newtheorem{conjecture}[proposition]{Conjecture}
\newtheorem{proposition-short}{Proposition} % a proposition in a short work (i.e., a work with no sections..) numbering is 1,2,3,...
\newtheorem{theorem-short}{Theorem} % a theorem in a short work (i.e., a work with no sections..) numbering is 1,2,3,...

\theoremstyle{remark}
\newtheorem{definition}{Definition} %[chapter] 
\newtheorem{example}{Example}
\newtheorem{examples}{Examples}
\newtheorem{remark}{Remark}

\newcommand{\thm}{\begin{theorem}}
\newcommand{\Thm}{\end{theorem}}
\newcommand{\thma}{\begin{theorem*}}
\newcommand{\Thma}{\end{theorem*}}

\newcommand{\conj}{\begin{conjecture}}
\newcommand{\Conj}{\end{conjecture}}
\newcommand{\prf}{\begin{proof}}
\newcommand{\Prf}{\end{proof}}
\newcommand{\dfn}{\begin{definition}}
\newcommand{\Dfn}{\end{definition}}
\newcommand{\prop}{\begin{proposition}}
\newcommand{\Prop}{\end{proposition}}
\newcommand{\cor}{\begin{corollary}}
\newcommand{\Cor}{\end{corollary}}
\newcommand{\cora}{\begin{corollary*}} % note: postfix asterisks "-a"
\newcommand{\Cora}{\end{corollary*}}
\newcommand{\lem}{\begin{lemma}}
\newcommand{\Lem}{\end{lemma}}
\newcommand{\lema}{\begin{lemma*}}
\newcommand{\Lema}{\end{lemma*}}
\newcommand{\rmk}{\begin{remark}}
\newcommand{\Rmk}{\end{remark}}
\newcommand{\exa}{\begin{example}}
\newcommand{\Exa}{\end{example}}
\newcommand{\exas}{\begin{examples}}
\newcommand{\Exas}{\end{examples}}
\newcommand{\meth}{\begin{method}}
\newcommand{\Meth}{\end{method}}
\newcommand{\bexa}{\begin{barredexample}}
\newcommand{\bExa}{\end{barredexample}}

\newcommand{\set}[1]{\{#1\}}% explicit set (curly braces)
\newcommand{\nll}{\text{\upshape{\O}}}% null
\newcommand{\nin}{\notin} % nin (is a singleton not contained in) 
\newcommand{\lies}{\subset} % lies (belongs to, contained in)
\newcommand{\settl}[2]{\left({}^{#1} \, #2 \right)} % set, settl notation 
\newcommand{\eand}{\text{ and }} % English and
\newcommand{\eimplies}{\text{ implies }} % English implies
\newcommand{\eif}{\text{ if }} % English if
\newcommand{\ethen}{\text{ then }} % English then
\newcommand{\id}{\text{id}} % identity morphism
\newcommand{\of}{\circ} % ordinary composition
\DeclareMathOperator{\dom}{dom} % domain
\DeclareMathOperator{\cod}{cod} % codomain
\newcommand{\iso}{\cong} % isomorphic (equals sign with tilde on top)
\DeclareMathOperator{\Ob}{Ob} % objects of a kind/cat
\DeclareMathOperator{\Sb}{Sb} % subjects of a kind
\newcommand{\dleq}{\sqsubseteq}

\newcommand{\vertof}{\mathbin{\raisebox{0.2ex}{$\scriptscriptstyle{\Delta}$}}} % vertical operation in Cat[2] "vertical of"
\newcommand{\trut}{\top} % true
\newcommand{\threeline}{\equiv} % three line equality 
\newcommand{\threelines}{\equiv} % three line equality 
\newcommand{\bij}{\overset{\text{bij}}{=}} % bijection
\newcommand{\arter}{ter} % terminal arrow
\newcommand{\name}[1]{\ulcorner #1 \urcorner} % name edge ---> term
\DeclareMathOperator{\ty}{ty} % type of _
\DeclareMathOperator{\FV}{FV} % free variables of _
\DeclareMathOperator{\CAP}{CAP} % captured variables of _
\DeclareMathOperator{\VAR}{VAR} % all variables appearing in _
\DeclareMathOperator{\deflate}{deflate} % ICCC and CCC
\newcommand{\prelambdaCalc}{{\bf pre}\text{-$\lambda$-}{\bf Calc}} % pre-generalized typed lambda calculus category
\newcommand{\lambdaCalc}{\text{$\lambda$-}{\bf Calc}} % generalized typed lambda calculus category
\DeclareMathOperator{\height}{height} % height
\DeclareMathOperator{\source}{{\bf s}} % source
\DeclareMathOperator{\target}{{\bf t}} % target
\newcommand{\Mu}{\text{M}} % capital mu

\usepackage{scalerel,stackengine}
\stackMath
\newcommand\verywidehat[1]{%
\savestack{\tmpbox}{\stretchto{%
  \scaleto{%
    \scalerel*[\widthof{\ensuremath{#1}}]{\kern-.6pt\bigwedge\kern-.6pt}%
    {\rule[-\textheight/2]{1ex}{\textheight}}%
  }{\textheight}% 
}{0.5ex}}%
\stackon[1pt]{#1}{\tmpbox}%
}

\begin{document}
\title{A Generalization of the Curry-Howard Correspondence}
\author{
Lucius Schoenbaum
}
\date{\today}

\begin{abstract}
We present a variant of the calculus of deductive systems developed in \cite{LaK1c,LaK2}, 
and give a generalization of the Curry-Howard-Lambek theorem giving an equivalence between the category of typed lambda-calculi and the %
category of cartesian closed categories and exponential-preserving morphisms that leverages the theory of generalized categories \cite{ScMd}.  %
We discuss potential applications and extensions. %
\end{abstract} 

\maketitle

\section{Introduction}\label{s.i}

In a series of papers \cite{LaK1c,LaK2}, Lambek developed an extension of the Curry-Howard correspondence \cite{Howard1969} to the domain of categorical logic. %
Lambek's extension has since become a cornerstone of programming language theory, particularly in the functional programming paradigm. It has also been influential in logic. % 
This paper %
is devoted to a generalization of the Curry-Howard-Lambek correspondence which makes use of the tools provided by generalized categories. %
Those who agree with Philip Wadler \cite{WaR5} that, as a general rule, semantics should guide development in logic and programming language theory may take interest in this product of a generalization on the semantic side. %
Those with a pure interest in category theory might note some features of our approach, for example, we show (section \ref{s.cat}) that using the framework of generalized categories, a cartesian functor between cartesian closed categories may be ``promoted'' to a cartesian closed functor. To the best of our knowledge this construction is at least somewhat new. 

Lambek in his work makes extensive use of deductive systems \cite{LaK1c}. %
A short discussion of the intuition for this notion (which may be unfamiliar) affords the opportunity to provide some intuition for the notion of generalized category. However, the reader is free to ignore this discussion if he or she wishes; nothing in the main body of the paper depends on it. %
A deductive system is just enough machinery to allow the question: from a given point $a$ of the deductive system $\sA$, can I travel to another point $b \in \sA$ via a valid path? %
A conceptual picture of this is the following. Suppose that there is a system of goods $\sA_0$. %
The edges of $\sA$ are certificates (issued, say perhaps, by different governing bodies) that say that a good $a \in \sA_0$ may be exchanged for another good $b \in \sA_0$. %
It is accepted that a good is always exchangeable for itself. %
Now let's suppose that such certificates themselves may be exchanged, % 
but that this requires that one has a higher-level certificate for this higher-level trade. %
If we imagine a certain impetus exists among those we imagine making the exchanges, we can expect that there will next arise trading for these certificates as well. % 
Let us make two simple observations:
\begin{enumerate}
	\item The resulting deductive system is not necessarily {\em cellular}, in the sense that the economy is liberalized to the extent that certificates may be good for exchange of different {\em kinds} of goods and certificates. For example, a certificate may be for a good, in return for a certificate good for a certificate in return for a good.
	\item There need not be, in the abstract, any {\em goods} at all. The system could be one of certificates for certificates for certificates, and so on. This observation may be utilized to clean up the abstract formalism: a system with no atoms is conceptually simpler and the easiest one to work with while developing elementary principles. 
\end{enumerate}
These two observations suggest, via the intuition, a generalization of category theory that we outline in section \ref{s.gencat}. 

Some work during intermediate stages is necessary in order to accomplish our aim. %
Under the Curry-Howard usual correspondence, types are interpreted as propositions which are true only when they are inhabited by a term. %
It is based on the types-as-targets view of categorical semantics, which limits the applicability of generalized categories to type theory. %
If we consider the alternative {\em types-as-paths} view, in which a proposition depends on both a source and a target, we find a calculus that is not only amenable to the generalized setting, but also fits well with the Lambek equational theory of cartesian closed categories \cite{LaK2}. %
The types-as-path view is motivated by the notion that a type is like the blueprint of a bridge between two points, or (in the logical intuition), a {\em conjecture}. %
Using the intuition from programs, on the other hand, the type-as-path is an {\em approximation} or {\em abstraction} of any choice of concrete transformation between two different kinds of data. This supports our approach, since this is how types are often viewed in applications, see for example \cite{Pierce1}. %
The types, which we write $a \vdash b$, when viewed categorically, assume the role of exponential objects. %
We are able to give this description a precise formal treatment by combining (1) the contributions of Lambek and (2) the framework of generalized category theory. 

In section \ref{s.gencat}, we introduce generalized categories \cite{ScMd}. In section \ref{s.th}, we develop ideal cartesian closed  categories, the notion we take of cartesian closed category in the generalized setting. These come equipped with an ideal of types, in the sense discussed above. In section \ref{s.poly}, we introduce polynomial categories, by closely following Lambek \cite{LaK2}, and in section \ref{s.lam} we define a notion of generalized type theoretic system (lambda calculus) corresponding to the semantics we have introduced, and verify that the anticipated equivalence holds. %
In all that we have done we have closely followed %
the well-established work of Lambek and others. However, our work lays the foundation for many possible avenues for further development in areas such as proof theory, programming language semantics, topos theory, and homotopy type theory. We discuss some topics for future work in section \ref{s.conc}. % 

\section{Generalized Categories}\label{s.gencat}

{\em Preliminaries.} %
We use notation $\source(f),\target(f)$, $\dom(f),\cod(f)$, and $\bar f, \hat f$, more or less interchangeably, to denote the source and target of an element of a generalized category. The lattermost notation may be used when it improves readability of formulas. %
We write composition $G \of F := (f \mapsto G(F(f)))$ and in general, for mappings $F$ and $G$ with common domain and codomain (in which concatenation is meaningful) we define the operation
$$G \vertof F := (f \mapsto G(f)F(f)),$$
the standard vertical composition operation \cite{MacCW}. %
In any context where it is meaningful, we use the standard arrow notation $f:a \to b$ to mean that an element $f$ is given, the source of $f$ is $a$, and the target of $f$ is $b$. %
The notation $\downarrow$ 
indicates that all composed pairs of elements in the expression or relation are in fact composable pairs. 

We recall the following from \cite{ScMd}. %
We restrict our focus to the sharp case. %

\dfn\label{d.gencat}
A {\em generalized category} is a structure $(\sC, \dleq, \source,\target, \cdot)$ where $\sC$ is a set, $\dleq$ is a relation on $\sC$, $\source$ and $\target$ are mappings $\sC \to \sC$, and $(\cdot)$ is a partially defined mapping $\sC \times \sC \to \sC$, denoted $a \cdot b$ or $ab$. These are required to satisfy
\begin{enumerate}
	\item $(\sC, \dleq)$ is a partially ordered set, % 
			\label{ax.gencat-po}
	\item $ab$ $\downarrow$ if and only if $\source(a) \dleq \target(b)$. \label{ax.gencat-po-comp}
	\item If $(ab)c$ $\downarrow$ or $a(bc)$ $\downarrow$ then $(ab)c = a(bc)$. \label{ax.gencat-assoc}
	\item If $ab$ $\downarrow$ then $\source(ab) = \source(b)$ and $\target(ab) = \target(a)$. \label{ax.gencat-comp-st}
	\item (Element-Identity) For all $a \in \sC$, there exists $b \in \sC$ such that \label{ax.gencat-element-id}
		\begin{enumerate}
			\item $\source(b) = \target(b) = a$,
			\item if $bc$ $\downarrow$ then $bc = c$,
			\item if $cb$ $\downarrow$ then $cb = c$,
		\end{enumerate}
	\item (Object-Identity) Let $a \in \sC$ and $\source(a) = \target(a) = a$. Then \label{ax.gencat-object-id}
		\begin{enumerate}
			\item if $ba$ $\downarrow$ then $ba = b$. 
			\item If $ab$ $\downarrow$ then $ab = b$.
		\end{enumerate}
The element $c$ of axiom (\ref{ax.gencat-element-id}) %
is unique, and is denoted $1_a$ or $\id_a$, and called the {\em identity} on $a$. %
	\item (Order Congruences)
		\begin{enumerate}
			\item If $a \dleq b$ then $\source(a) \dleq \source(b)$ and $\target(a) \dleq \target(b)$. \label{ax.gencat-order1}
			\item $a \dleq b$ and $c \dleq d$ and $ac,bd$ $\downarrow$ implies $ac \dleq bd.$ \label{ax.gencat-order2}
			\item $a \dleq b$ implies $1_a \dleq 1_b$. \label{ax.order3}
		\end{enumerate}
\end{enumerate}
\Dfn

A generalized category $\sC$ is {\em 1-dimensional} or a {\em one-category} if $st = t, tt = t, ss = s, ts = s$ in $\sC$, where $s$ and $t$ are the source and target operators. There is a bijection between 1-dimensional categories and ordinary categories given by embedding objects in the set of arrows via $X \mapsto \id_X$. 
An {\em element} $f \in \sC$ is an element $f$ of the underlying set $\sC$. 
An {\em object} $a$ in $\sC$ is an element $a$ of $\sC$ such that $\source(a) = \target(a) = a$. 
We write $\Ob(\sC)$ for the set of objects. 

\dfn\label{d.functor}
Let $\sC, \sD$ be generalized categories. A {\em functor} from $\sC$ to $\sD$ is a structure-preserving map, that is, a mapping $F: \sC \to \sD$ satisfying:
\begin{enumerate}
	\item $\rule{0ex}{2.8ex} F(gf) = F(g) F(f)$ %
	\item $\rule{0ex}{2.8ex} F(\bar f) = \overline{F(f)}$
	\item $\rule{0ex}{2.8ex} F(\hat f) = \widehat{F(f)}$
\end{enumerate}
\Dfn

\dfn
Let $\sC$ and $\sD$ be generalized categories, and let $F,G:\sC \to \sD$ be functors. A {\em natural transformation} from $\sC$ to $\sD$ is a mapping $\theta:\sC \to \sD$ such that for every $f,g \in \sC$, 
\[
	\theta(\hat f)F(f) = G(f)\theta(\bar f) \,\, \downarrow
\]
\Dfn

The class of all generalized categories (in the fixed universe), functors between generalized categories, and 
natural transformations between functors form a strict 2-category. %

\dfn\label{d.adjunction}
Let $\sC$ and $\sD$ be generalized categories. An %
{\em adjunction} %
$(F,G,\eta,\epsilon)$ is a pair of functors 
\[
\raisebox{-0.5\height}{\includegraphics{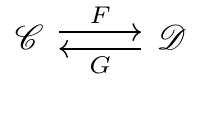}} 
\]

\vspace{-15pt}
\noindent together with %
natural transformations
$$
\eta: \id_\sC \to G\of F, \quad \epsilon: F \of G \to \id_\sD,
$$
satisfying the identities
\begin{align}\label{eq.triid}
(G \of \epsilon) \vertof (\eta \of G) &= 1_G, \\
(\epsilon \of F) \vertof (F \of \eta) &= 1_F,
\end{align}
where $1_F$ is the mapping $f \mapsto 1_{F(f)}$. 
\Dfn

Equivalently, given such an $F$ and $G$, 
for every $f$ in $\sC$ and $g$ in $\sD$, there is a 
bijection of sets
$$\hom(F(f), g) \iso \hom(f, G(g)),$$
that is natural in $f$ and $g$. This means that if $\phi$ is the bijection, then $\phi$ satisfies
\[
u \cdot F(v): F(f) \to g \eimplies \phi(u \cdot F(v)) = \phi(u) \cdot v,
\]
\[
v' \cdot v: F(f) \to g \eimplies \phi(v' \cdot v) = G(v') \cdot \phi(v).
\]

Many notions of category theory 
\cite{MacCW,BaWe1} have been carried over to the generalized setting, though some constructions require more effort than others. For example:

\dfn\label{d.monad}
Let $\sC$ be a generalized category. A {\em monad} on $\sC$ is a structure $(T,\eta,\mu)$, where $T: \sC \to \sC$ is a functor, and $\eta$ and $\mu$ are (order-preserving) natural transformations $\id_\sC \to T$ and $T^2 \to T$, respectively, such that the following hold:
\begin{enumerate}
	\item $\mu \vertof (T \of \mu) = \mu \vertof (\mu \of T)$ \label{ax.monmagic1}
	\item $\mu \vertof (T \of \eta) = \mu \vertof (\eta \of T) = 1_T$, \label{ax.monmagic2}
\end{enumerate}
where $1_T$ denotes the mapping $f \mapsto 1_{T(f)}$. %
for all $x,y$ in $\sC$, 
\Dfn

The relation between monads and triples and the Tripleability Theorems were studied in \cite{ScMd}.

%

%

%

% iccc
\section[Gen. Deductive Systems and Ideal Cartesian Closed Categories]{Generalized Deductive Systems and Ideal Cartesian Closed Categories}\label{s.th}

\subsection{Generalized Deductive Systems and Generalized Graphs}\label{ss.gendedsys}

\dfn\label{d.gengraph}
A {\em generalized graph} is a triple $(\sA, s, t)$, where $\sA$ is a set, %
$s, t$ are maps $\sA \to \sA$. %

A {\em morphism} $\Phi: \sA \to \sB$ of generalized graphs is a mapping $\Phi$ from $\sA$ to $\sB$ such that 
\begin{enumerate}
	\item $\Phi(s(a)) = s(\Phi(a))$
	\item $\Phi(t(a)) = t(\Phi(a))$
\end{enumerate}
This gives a category $\bf{Graph}$ of generalized graphs.
\Dfn

An element of $\sA$ is (synonymously) an {\em edge}. %
An {\em object} in a generalized graph is an element $a \in \sA$ such that $sa = ta = a$. %
We say that a {\em subject} in a generalized graph is an element $a \in \sA$ such that there is an element $f \in \sA$ such that either $sf = a$ or $tf = a$. %
We write $\Ob(\sA)$, $\Sb(\sA)$ for the set of objects and subjects of $\sA$, respectively. %
We say that generalized graph is {\em 1-dimensional} if 
$$ss = s \text{ and } tt = t.$$
Ordinary graphs correspond bijectively with 1-dimensional generalized graphs. %

Recall that in an algebraic system $(A, f)$ in which $A$ is a carrier set where equality $(=)$ is defined and a unary operation $f$ is defined (a mapping $A \to A$), we say that $f$ is {\em substitutive} if for all $a,b \in A$
$$a = b \eimplies fa = fb.$$
(The word {\em congruence} also arises frequently in connection with this property.) %
The source and target operations in a generalized graph are not assumed to be substitutive. %
(In fact, there is no notion of equality defined in the language of generalized graphs until we come to Definition \ref{d.icat}.) %
This comes with the advantage that we can apply inductive pattern-matching in proofs about elements in a generalized graph (and we may even do so constructively, if they are finitely generated in some finite language), %
though yet another hypothesis is needed if these patterns matchings are to be exhaustive in $\sA$. (Such a hypothesis will apply to polynomials in section \ref{s.poly}.)

\dfn\label{d.dedsys}
A {\em generalized typed deductive system} or simply a {\em generalized deductive system} is a structure 
$$(\sA, s, t, \cdot, \vdash, \sV),$$ 
where $(\sA, s, t)$ is a generalized graph, %
($\cdot$) is a partially-defined operation $\sA \times \sA \to \sA$ on $\sA$,
$(\vdash)$ is an operation $\sA \times \sA \to \sA$, %
and $\sV$ is a subset of $\sA$, satisfying 
\begin{enumerate}
	\item for all $a,b \in \sA$, $b \cdot a \text{ is defined iff } ta = sb$
	\item $s(ab) = s(b)$ and $t(ab) = t(a)$
	\item $s(a \vdash b) = a$ and $t(a \vdash b) = b$. \label{ax.vdashst}
	\item for every $a \in \sA$, $a \vdash a \in \sV$. \label{ax.identityvalid}
	\item if $a,b \in \sV$, and $a \cdot b$ $\downarrow$, then $a \cdot b \in \sV$. \label{ax.compositionvalid}
	\item for every $a,b \in \sA$, if there exists $u \in \sV$ with $\bar u = a$ and $\hat u = b$, then $a \vdash b \in \sV$. \label{ax.inhabitance}
\end{enumerate}
A morphism of generalized typed deductive systems $\phi: \sA \to \sB$ is a morphism of generalized graphs satisfying
\begin{enumerate}
	\item $\phi(a \cdot b) = \phi(a) \cdot \phi(b),$
	\item $\phi(a \vdash b) = \phi(a) \vdash \phi(b),$
	\item if $a \in \sV$, then $\phi(a) \in \sV$.
\end{enumerate}
This gives a category ${\bf DedSys}$ of generalized (typed) deductive systems.
\Dfn

Since we can now compose edges, we shall refer elements $a$ of a deductive system $\sA$ as edges or {\em paths} (there is no actual distinction between the two terms, except in case products in $(\cdot)$ are freely generated on a basis in $\sA$.) %
The elements of $\sV$ may be thought of as {\em valid paths} of $\sA$. %
In the set of edges going from $a$ to $b$, the unique edge $a \vdash b$ is called the {\em type} with source $a$ and target $b$. %
We may use the notation $a \dashv b$ interchangeably to denote $b \vdash a$, thus $a\dashv b \threeline b \vdash a$. 
Finally, when using axiom (\ref{ax.inhabitance}) we call $u$ a {\em witness} and say that the type $a \vdash b$ is {\em inhabited} if there is found such a $u$. %
We may write $1_a$ in place of $a \vdash a$. %

In our work it is possible to ignore the role of $\sV$, but its presence suggests generalizations of the calculus, for example $\sV$ %
might be useful in a model of concurrency, or be impacted by modal operators.

\dfn\label{d.pidedsys}
An {\em positive intuitionistic generalized deductive system} is a generalized deductive system 
$$(\sA, s, t, \cdot, \vdash, \sV)$$ 
equipped with the additional structure
$$(\trut, \wedge, \langle,\rangle, ()^*)$$
consisting of:
\begin{enumerate}
	\item A distinguished element $\trut \in \sA$,
	\item A mapping $\wedge: \sA \times \sA \to \sA$,
	\item A partially defined mapping $\langle,\rangle : \sA \times \sA \to \sA$,
	\item A partially defined mapping $()^* : \sA \to \sA$
\end{enumerate}
subject to the following axioms,
\begin{enumerate}
	\item $\langle a,b \rangle$ is defined if and only if the source of $a$ and $b$ are identical.
	\item $a^*$ is defined if and only if the source of $a$ is the wedge of two subjects in $\sA$.
\end{enumerate}
the following source and target conditions:
\begin{enumerate}
	\item $\hat\trut = \bar\trut = \trut$, %
	\item $s(a \wedge b) = s(a) \wedge s(b)$ and $t(a \wedge b) = t(a) \wedge t(b)$, 
	\item $s(\langle a , b \rangle) = s(a)$ and $t(\langle a, b \rangle) = t(a) \wedge t(b)$
	\item $s(a^*) = proj_1(s(a))$ and $t(a^*) = proj_2 (s(a)) \vdash t(a)$, where $proj_1$ and $proj_2$ are the projections on wedge ($\wedge$) products.
\end{enumerate}
and the following {\em rules}, or {\em validities:} For all $a,b \in \sV$,
\begin{enumerate}
	\item $\trut \in \sV$,
	\item $a \wedge b$ is valid if $a$ and $b$ are valid,
	\item $\langle a, b \rangle$ is valid if $a$ and $b$ are valid,
	\item $a^*$ is valid if $a$ is valid,
	\item for every pair of subjects $a,b \in \sA$, the following types are valid: \label{ru.validatoms}
		\begin{enumerate}
			\item $\trut \vdash a$
			\item $a \wedge b \vdash a$ 
			\item $a \wedge b \vdash b$ % 
			\item $(a \vdash b) \wedge a \vdash b$ 
		\end{enumerate}
\end{enumerate}
A morphism $f: \sA \to \sB$ of positive intuitionistic generalized deductive systems is a morphism of generalized deductive systems satisfying 
\begin{enumerate}
	\item $F(\trut) = \trut,$
	\item $F(a \wedge b) = F(a) \wedge F(b),$
	\item $F(\langle f, g \rangle ) = \langle F(f), F(g) \rangle,$
	\item $F(f^*) = F(f)^*.$
\end{enumerate}
This gives a category ${\bf p.i.DedSys}$ of positive intuitionistic generalized deductive systems. 
\Dfn

In order to form complex expressions out of simple ones, it is convenient to have names for individual elements of $\sA$. % 
For example, we choose (applying rule \ref{ru.validatoms}) valid elements of $\sA$
	$$ter_a : a \to \trut$$
	$$\pi_{a,b} : a \wedge b \to a$$
	$$\pi_{a,b} : a \wedge b \to b$$
	$$\epsilon_{a,b} : (a \vdash b) \wedge a \to b$$
Note that these elements may themselves be types, even though we usually think of the types as valid due to the existence of a witness and use of axiom \ref{ax.inhabitance} of Definition \ref{d.dedsys}. %
By {\em term} (or {\em global element}) of a deductive system we refer to any element of a deductive system whose source is $\trut$. 

One may use the deductive system to show the validity of Heyting's axioms for intuitionistic logic (those that do not contain the $\lor$ and $\bot$ connectives), showing that any type that may be interpreted as a valid proposition of intuitionistic logic has a witness. The following types, for example, are inhabited.
\begin{enumerate}
	\item $a \vdash a \wedge a$
	\item $a \vdash a \wedge \trut$
	\item $((a \wedge b) \vdash c) \vdash (a \vdash (b \vdash c))$
	\item $(a \vdash (b \wedge c)) \dashv \vdash ((a \vdash b) \wedge (a \vdash c))$
\end{enumerate}
where $\dashv \vdash$ denotes that the type is bi-inhabited (or there is a valid path going in either direction). %

\subsection{Categories Equationally Defined}\label{s.cat}
Lambek \cite{LaK1c,LaSc1} observed that categories are obtained from deductive systems via a set of equational axioms. %
In this section we will develop Lambek's formalization in the setting of generalized categories. %
It is clear that any ordinary category (or generalized category) can be made into a ``typed deductive'' category. %
Simply take all arrows to be valid and introduce $\vdash$ as a free operation %
Observe that if composition $(\cdot)$ is viewed as multiplication and $\wedge$ is viewed as an additive product on the subjects of $\sC$, the set of elements of the form $a \vdash b$ behaves like a (ring-theoretic) ideal in the category. Thus if we are thinking of a category, we may think next of introducing an ``ideal of types'' to the category. %
This demands we introduce a further technicality, a set of constants. %

\dfn\label{d.icat}
A {\em ideal category} or {\em ideal generalized category} is a structure $(\sC, \vdash, \sV)$ consisting of a generalized category $\sC$ (section \ref{s.gencat}), %
a distinguished subset of elements $\sV \lies \sC$, %
a distinguished subset of elements $\sK \lies \sC$,
and an operation $\vdash: \sC \times \sC \to \sC$ such that 
\begin{enumerate}
	\item $\verywidehat{f \vdash g} = f,$
	\item $\overline{f \vdash g} = g.$
	\item $f \cdot (\bar f \vdash g) = \hat f \vdash g,$ unless $\bar f = g$, in which case $f \cdot (\bar f \vdash g) = f$, or unless $f \in \sK$ or $\bar f \vdash g \in \sK$. \label{ax.vdash1}
	\item $(f \vdash \hat g) \cdot g = f \vdash \bar g,$ unless $f = \hat g$, in which case $(f \vdash \hat g) \cdot g = g$, \label{ax.vdash2}
	\item if $g \cdot f$ $\downarrow$, and $g,f \in \sV$, then $g \cdot f \in \sV$,
	\item $f \vdash f \in \sV$ for all $f \in \sC$,
	\item (witnesses) $u \in \sV$ implies $\hat u \vdash \bar u \in \sV$.
	\item $\vdash$ is substitutive (Section \ref{ss.gendedsys}) in both arguments. 
\end{enumerate}
A functor $F: \sC \to \sD$ between generalized ideal categories is an ordinary functor (section \ref{s.gencat}) which preserves validity and $\vdash$:
\begin{enumerate}
	\item $f \in \sV_\sC \eimplies F(f) \in \sV_\sD,$
	\item $F(f \vdash g) = F(f) \vdash F(g)$. \label{ax.deductivefunctor2}
\end{enumerate}
This defines a category ${\bf IdealCat}$. 
\Dfn

By axioms \ref{ax.vdash1} and \ref{ax.vdash2}, for $f \in \sC$, $f \vdash f$ is the identity of $f$, which may be denoted $1_f$. %
In particular, all elements (including identities) of an ideal category have identities. 
The identities, types, and constants figuring here will arise again in Section \ref{s.lam}, where we encounter the symbols $\settl{x}{x}$ and $\settl{x}{y}$. %

\dfn\label{d.iccc}
An {\em ideal cartesian closed category} is a ideal category with identities $\sC$ that is equipped with a structure
$$(\trut, \wedge, \langle,\rangle, ()^*)$$
where
\begin{enumerate}
	\item $\trut$ is a distinguished valid element in $\sC$,
	\item $\wedge$ is an operation $\sC \times \sC \to \sC$,
	\item $\langle, \rangle$ is a partially defined operation $\sC \times \sC \to \sC$
	\item $()^*$ is a partially defined operation $\sC \to \sC$
\end{enumerate}
which satisfies the conditions:
\begin{enumerate}
	\item $\trut \in \sK$, and $\sK$ is closed under $\wedge, \langle,\rangle, $ and $()^*$,
	\item the structure
		$$(s,t,\sV,\vdash, \cdot, \trut, \wedge, \langle,\rangle, ()^*)$$
		defines a positive intuitionistic deductive system on $\sC$.
	\item for all $a \in \sC$, if $f: a \to \trut$ then $f = (a \vdash \trut)$. \label{ax.terminalar}
	\item For every pair $(a,b)$ of subjects of $\sC$, there exists a good pair $(\pi, \pi')$ for $(a,b)$. 
	\item For every good pair $(\pi, \pi')$ for any pair of subjects $(a,b)$, there is a good evaluation $\epsilon = \epsilon_{\pi, \pi'}$ for $(\pi, \pi')$. 
\end{enumerate}
Here, if $(a,b)$ is a pair of subjects of $\sC$, then a pair $(\pi, \pi')$ of elements of $\sC$ are a {\em good pair for $(a,b)$} if
			\begin{enumerate}
				\item $\pi$ and $\pi'$ are valid,
				\item $\pi: a \wedge b \to a$, and $\pi': a \wedge b \to b$,
				\item if $\pi \langle f, g \rangle$ $\downarrow$ then $\pi \langle f, g \rangle = f,$
				\item if $\pi' \langle f, g \rangle$ $\downarrow$ then $\pi' \langle f, g \rangle = g,$
				\item if $\langle \pi f, \pi' f \rangle$ $\downarrow$ then $\langle \pi f, \pi' f \rangle = f$, % 
				\item if $f \cdot \pi$ and $g \cdot \pi'$ $\downarrow$, then $\langle f \cdot \pi, g \cdot \pi' \rangle = f \wedge g$. 
			\end{enumerate}
and a {\em good evaluation} for a good pair $(\pi, \pi')$ for a pair of subjects $(a,b)$ is an element $\epsilon = \epsilon_{\pi, \pi'}$ of $\sC$ that satisfies, for every $c \in \sC$ and every good pair $(\pi_{c,a}, \pi'_{c,a})$ for $(c,a)$,
			\begin{enumerate}
				\item $\epsilon$ is valid,
				\item $\epsilon: (a \vdash b) \wedge a \to b$,
				\item if $\epsilon \cdot \langle f^* \cdot \pi_{c,a}, \pi'_{c,a} \rangle$ $\downarrow$ then $\epsilon \cdot \langle f^* \cdot \pi_{c,a}, \pi'_{c,a} \rangle = f,$ 
				\item if $(\epsilon \cdot \langle f \cdot \pi_{c,a}, \pi'_{c,a} \rangle )^*$ $\downarrow$ then $(\epsilon \cdot \langle f \cdot \pi_{c,a}, \pi'_{c,a} \rangle )^* = f$. % 
			\end{enumerate}
A morphism $F: \sC \to \sD$ between ideal cartesian closed categories $\sC$ and $\sD$ is a functor of ideal categories satisfying
\begin{enumerate}
	\item $F(\trut) = \trut$,
	\item $F(a \wedge b) = F(a) \wedge F(b),$
	\item $F(\langle a , b \rangle) = \langle F(a), F(b) \rangle,$
	\item $F$ sends a good pair in $\sC$ to a good pair in $\sD$. %
\end{enumerate}
Thus we have a category ${\bf ICCC}$ of ideal cartesian closed categories.
\Dfn

Axiom \ref{ax.terminalar} is relevant when the possibility exists that the element $a \vdash \trut$ might be a constant. 
We continue to use the notation of deductive systems in a category $\sC$. Note that many authors write $\times$ for the binary product, which we continue to denote $\wedge$, and $1$ for the terminal object, which we continue to denote $\trut$. This seems appropriate as we will never stray far from the point of view provided by deductive systems and the lambda calculus. 

Note that morphisms of ideal cartesian closed categories are stronger maps than ordinary functors between categories that happen to be cartesian closed. For ordinary categories, these functors are sometimes called {\em cartesian functors}. %
It is easy to see that a good pair $(\pi, \pi')$ for a pair of subjects $(a,b)$ is unique if it exists. Hence a good evaluation $\epsilon \threelines \epsilon_{\pi, \pi'}$ depends only on $(a,b)$ and may be denoted $\epsilon_{a,b}$. %, 
Similarly, we often write $\pi \threelines \pi_{a,b}$ and $\pi' \threelines \pi'_{a,b}$. It follows that $F(\pi_{a,b}) = \pi_{F(a), F(b)}$, and similarly for $\pi'$.

\prop\label{p.basiciccc}
The following hold in ideal cartesian closed categories:
\begin{enumerate}
	\item $\langle f, g \rangle \cdot h = \langle f \cdot h , g \cdot h \rangle$
	\item $1_a \wedge 1_b = 1_{a \wedge b}$
	\item $(f \wedge g) \cdot (f' \wedge g') = (f \cdot f') \wedge (g \cdot g')$
	\item $\epsilon_{a,b}^* = 1_{a \vdash b}$
	\item $f^* \cdot g$ $\downarrow$ implies $f^* \cdot g = (f \cdot \langle g \cdot \pi, \pi' \rangle)^*$, where $(\pi, \pi')$ is the obvious good pair.
\end{enumerate}
\Prop

\prop
A morphism $F: \sC \to \sD$ between ideal cartesian closed categories preserves the evaluation $\epsilon$ and adjoint operation $()^*$. %
\Prop
\prf
By functoriality, we have $F((f: a \wedge b \to c)^*) = F(f^*) : F(a) \to F(b \vdash c) = F(f^*) : F(a) \to (F(b) \vdash F(c))$. But this latter expression is $F(f)^*$, so $F(f^*) = F(f)^*$. %
It follows that evaluations $\epsilon$ are also preserved. Indeed, if $(a,b)$ are chosen and $(\pi, \pi')$ is a good pair for $(a,b)$, then choose a good evaluation $\epsilon = \epsilon_{a,b}$ for $(\pi, \pi')$. %
Then 
$$F(\epsilon)^* = F(\epsilon^*) = F(1_{a \vdash b}) = 1_{F(a) \vdash F(b)}.$$
Hence 
\begin{align*}
\epsilon_{F(a), F(b)} 	&= \epsilon_{F(a), F(b)} \cdot \langle 1_{F(a) \vdash F(b)} \pi_{F(a) \vdash F(b) ,a}, \pi'_{F(a) \vdash F(b),a} \rangle  \\
				&= \epsilon_{F(a), F(b)} \cdot \langle F(\epsilon_{a,b})^* \pi_{c,a}, \pi'_{c,a}\rangle  \\
				&= F(\epsilon_{a,b}),
\end{align*}
by the good evaluation properties of $\epsilon_{a,b}$. 
\Prf

Next we present a few ways to produce ideal cartesian closed categories:

\prop\label{p.cattoicat}
There is an (in general, nonconstructive) functor from the category ${\bf CCC}$ of cartesian closed categories to the category {\bf ICCC}. 

\Prop
\prf
Let $F: \sC \to \sD$ be a functor in the category of cartesian closed categories (of the ordinary sort). We carry out the following construction on both $\sC$ and $\sD$; first take $\sC$. Take any new pair of identifiers $\vdash$ and $\wedge$. For each object $X$ of $\sC$, form, via recursion, the collections of triples
$$\sA_X = \set{(Y_1, Z_1, \vdash) \mid \text{ there exists } Y, Z \in \Ob(\sC) \text{ such that } X = Z^Y \eand Y_1 \in \sC_Y, Z_1 \in \sC_Z}$$
$$\sB_X = \set{(Y_1, Z_1, \wedge) \mid \text{ there exists } Y, Z \in \Ob(\sC) \text{ such that } X = Y \wedge Z \eand Y_1 \in \sC_Y, Z_1 \in \sC_Z}$$
$$\sC_X = \sA_X \cup \sB_X.$$
We take
$$\Ob(\tilde{\sC}) = \Ob(\sC) \cup \bigcup_{X \in \Ob(\sC)} \sC_X,$$
and for each $V \in \Ob(\tilde{\sC})$ we assume given from the construction of the $\sC_X$'s a function $\deflate(V)$ defined by
$$\deflate(V) = \begin{cases} V, & \eif V \in \sC, \\ Z^Y, &\eif V \in \sA_X \text{ for some $X$}, \\ Y \wedge Z, & \eif V \in \sB_X \text{ for some $X$}. \end{cases}$$
For every $U,V \in \tilde \sC$, define
$$\hom(U,V) := \hom(\deflate(U), \deflate(V)),$$
with composition and identities defined in the obvious way, in particular 
$$\deflate(f \cdot g) := \deflate(f) \cdot \deflate(g),$$
where $\deflate(f)$ for a morphism $f$ is defined in the obvious way analogous to $\deflate()$ on objects. 
The reader can now check that the symbols in Definition \ref{d.iccc} may be introduced and the axioms verified, and that we may extend $F$ to a functor $\tilde F: \tilde \sC \to \tilde \sD$ that satisfies the conditions of Definitions \ref{d.icat} and \ref{d.iccc}. 
\Prf

Another result that gives examples of ideal cartesian closed categories is:

\prop
Let $\sE$ be a generalized category of generalized presheaves over a generalized category $\sC$. Then $\sE$ is an ideal cartesian closed category. 
\Prop
\prf 
See  
\cite{ScMd}.
\Prf

The adjunction that holds in a cartesian closed category, because the mappings $- \times X$ and $-^X$ are no longer functors in the generalized setting. However, we do have:

\prop
\begin{enumerate}
	\item there is a bijection  
		$$\hom(c \wedge b, a) \bij \hom(c, a^b)$$
	\item there is a bijection
		$$\hom(a,b) \bij \hom(\trut, b^a)$$
\end{enumerate}
Let $a \iso b$ denote that there exists a pair of elements $f:a \to b$ and $g:b \to a$ such that $fg = 1_b$ and $gf = 1_a$. Then in an ideal cartesian closed category
\begin{enumerate}
	\item $(a \wedge b) \vdash c \iso (a \vdash b) \vdash c$
	\item $a \vdash (b \wedge c) \iso (a \vdash b) \wedge (a \vdash c)$
\end{enumerate}
\Prop
\prf
See \cite{LaSc1}. 
\Prf

Given $f:a \to b$ we write 
$$\name f $$ %
for the induced term $1 \to a \vdash b$, called (Lawvere's terminology) the {\em name} of $f$. 

Finally, we relate deductive systems to categories as follows:

\prop
Every deductive system $\sA$ on which there is defined an equivalence relation denoted $=$, and a distinguished subset $\sK$ of constants in $\sA$, with respect to which the following statements are satisfied:
\begin{enumerate}
	\item $f \cdot (\bar f \vdash g) = \hat f \vdash g,$ unless $\bar f = g$, in which case $f \cdot (\bar f \vdash g) = f$, unless $f$ is constant or $\bar f \vdash g$ is constant,
	\item $(f \vdash \hat g) \cdot g = f \vdash \bar g,$ unless $f = \hat g$, in which case $(f \vdash \hat g) \cdot g = g$, 
	\item $(hg)f = h(gf)$ for all composable $f,g,h \in \sA$,
	\item $a = b$ implies $s(a) = s(b)$,
	\item $a = b$ implies $t(a) = t(b)$,
	\item $a = b$ implies $ca = cb$ and $ac = bc$, for all composable $c$,
	\item $a = b$ implies $a \vdash c = b \vdash c$ and $c \vdash a = c \vdash b$, for all $c$ in $\sA$,
\end{enumerate}
is an ideal generalized category (in particular, a generalized category), taking $\sV$ to be the valid paths in $\sA$. 
\Prop
\prf
We check the axioms of Definition \ref{d.gencat} and see that they may be verified using axioms and rules of Definitions \ref{d.dedsys} and \ref{d.icat}. %
\Prf

The notion of a cartesian closed category cannot be extended to the generalized setting: the mapping $X \mapsto X \times Y$ is a functor only when $X$ is an object. %
Our approach is to allow %
the mapping on the other side, $Z \mapsto Z^Y$, to fail to be a functor as well. %
This is possible thanks to Lambek's formalization: %
We are able, by following Lambek, to derive a calculus of cartesian closed categories in the generalized setting, in spite of the weaker underlying structure. %

\section{Polynomials and Lambda-Calculi}\label{s.poly}

Adding variables to a deductive system with a positive intuitionistic structure reduces, by the Deduction Theorem (Theorem \ref{t.dedt}), validity of all paths to the validity of paths from a terminal object. Therefore the focus shifts from the space to the {\em polynomials over the space}, in the sense we now define. %

\subsection{Polynomials Systems and Polynomial Categories}\label{ss.poly}
The notion of indeterminate may be applied in this setting just as it may be applied in the setting of groups, rings, and fields. However, %
we must assign a source and target to each new indeterminate. It is convenient to let the source of every indeterminate be $1$, the (fixed choice of) terminal object. This does not mean we cannot substitute a variable with a different source for the indeterminate---substitution of, say, $a$ for $x$ in $\phi(x)$ is allowed whenever $x$ and $a$ have the same target; the source of $a$ is irrelevant. %
In this sense, it is more correct (but less convenient) to say that an indeterminate simply does not have a source. %
We denote an indeterminate over a deduction system $\sA$ by symbols $x, y, z, $ etc. %
For now, we require that the target of $x, y, \dots$ is in $\sA$. (In particular, it cannot itself be a polynomial). %
A more general system might allow indeterminates over polynomials and make use of the notion of {\em telescope} \cite{deBruijn1}, but we will have no need for this added generality. 

\dfn\label{d.poly}
Let $\sA$ be a positive intuitionistic deductive system. %
Let $x$ be an indeterminate with target $\hat x$ in $\sA$.
We write $\sA[x]$ for the positive intuitionistic deduction system freely % 
generated on the set $\sA \cup \set{x}$. %
This means that
\begin{enumerate}
	\item Operations on $\sA$ of Definition \ref{d.pidedsys} are extended from $\sA$ to $\sA[x]$ by free generation on expressions $\phi$ containing any instance of $x$: %
		$$\phi ::= f \,\mid\, x \,\mid\, %
		\phi \vdash \phi \,\mid\, \phi \wedge \phi \,\mid\, \langle \phi, \phi \rangle \,\mid\, \phi^* $$
	where $f$ can be any element of $\sA$, and $x$ is any indeterminate. Expressions so generated that do not contain any instance of $x$ are thrown out, and the set of all elements of $\sA$ is then added back in. 
	\item The valid elements of $\sA[x]$ are $x$, those of $\sA$, and those generated from $x$ and those of $\sA$ using the validities of Definition \ref{d.pidedsys}.
\end{enumerate}
There is an obvious embedding of $\sA$ in $\sA[x]$, via which we will usually view $\sA$ as a subset of  $\sA[x]$. 
\Dfn

We call elements of $\sA[x]$ synonymously {\em polynomials over $\sA$}. %
We write $\phi, \psi, \dots$ to denote polynomials in $\sA[x]$. %
We do not normally write the variable $x$ as in $\phi(x),$ etc. as many authors do, but this should not lead to any confusion as long as it is understood what may depend on $x$.
When we iterate to form $\sA[x][y]$, etc., we again require that the source and target of indeterminates be in $\sA$. Given indeterminates $x_1, x_2, \dots, x_n$, we denote by $\sA[x_1, \dots, x_n]$ or $\sA[\vec x]$ the iterated construction $(\dots((\sA[x_1])[x_2]) \dots [x_n])$. 

We could define a ``proof'' to be a valid path from the terminal object $\trut$ in a positive intuitionistic deductive system (say). Then we could ask what structure might allow us to ``discharge'' assumptions, as is done in natural deduction systems (see for example \cite{TrSc1}). %
To refine the question, one may consider a proof $\phi$ of $f \in \sA[x]$, for $f \in \sA$. 
This would be a path through the deductive system that is allowed to ``use'' the ``assumption'' $x$. %
In logic, the following result is, by long tradition, known as the Deduction Theorem. It is interpreted as an introduction rule when the construction of polynomials is interpreted as establishing a context. Note that polynomials do not necessarily have an element of $\sA$ as source and target, so the quantifiers on $a$ and $b$ are a significant part of the statement. (These ``higher'' polynomials arise in \cite{ScMd}.)

\thm\label{t.dedt}
Let $\sA$ be a positive intuitionistic deductive system. 
Then for all $a,b \in \sA$, $a \vdash b$ is valid in $\sA[x]$ if and only if $\hat x \wedge a \vdash b$ is valid in $\sA$. %
\Thm
\prf
The proof is just as in \cite{LaSc1}, except that we must add clauses for the operations $\wedge$ and $\vdash$. %
Note that several steps depend on the existence of identities on the subjects of $\sA$, as assumed in definition \ref{d.pidedsys}. 
First, % 
let $f$ be a valid path from $\hat x \wedge a$ to $b$ in $\sA$. Then since $\phi := \langle (x \cdot (a \vdash \trut) , 1_a \rangle $ is a valid path from $a$ to $\hat x \wedge a$ in $\sA[x]$, we obtain a witness $f \cdot \phi$ of the type $a \vdash b$ in $\sA[x]$, as desired. 

Now suppose $\phi$ is a valid path from $a$ to $b$ in $\sA[x]$. Suppose that for all polynomials in $x$ $\phi_<$ of length strictly less than $\phi$, there is a witness of $\hat x \wedge \overline{\phi_<} \vdash \widehat{\phi_<}$, denoted %
$$\kappa_x (\phi_<).$$
Now we proceed by cases:
\begin{enumerate}
	\item if $\phi \in \sA$, then $\phi \cdot \pi'_{\hat x, a}$ validates $\hat x \wedge a \vdash b$. \label{case.constant}
	\item if $\phi = x$, then $\pi_{\hat x, a}$ validates $\hat x \wedge a \vdash b$. 
	\item if $\phi = \psi \vdash \chi$ for some $\psi, \chi \in \sA[x]$, then 
			$a$ is identical to $\psi$ and $b$ is identical to $\chi$, hence this case reduces to case (\ref{case.constant}).
	\item if $\phi = \psi \cdot \chi$ for some $\psi, \chi \in \sA[x]$, then 
			$$\kappa_x \psi \cdot \langle \pi_{\hat x, a}, \kappa_x \chi \rangle$$
			is the desired witness. ($\chi \cdot \kappa_x \psi$ doesn't work, because $x$ is still not eliminated.)	
	\item if $\phi = \psi \wedge \chi$ for some $\psi, \chi \in \sA[x]$, then 
			$$\langle \kappa_x (\psi) \cdot \pi_{\bar{\psi}, \bar{\chi}}, \kappa_x (\chi) \cdot \pi'_{\bar{\psi}, \bar{\chi}} \rangle $$
			is the desired witness. (The alternative witness $\kappa_x \psi \wedge \chi_x \cdot \lambda$, where $\lambda$ is a munging factor, gives a definition of $\kappa_x$ under which one does not prove Theorem \ref{t.fCompT}.)
	\item if $\phi = \langle \psi, \chi \rangle$ for some $\psi, \chi \in \sA[x]$, then
			$$\langle \kappa_x \psi, \kappa_x \chi \rangle$$
			is the desired witness.
	\item if $\phi = \psi^*$ for some $\psi \in \sA[x]$, then
			$$(\kappa_x (\psi )\cdot \alpha)^*$$
			is the desired witness, where $\alpha$ is the associator.
\end{enumerate}
Proceeding by induction on the length of polynomials $\phi$ in $\sA[x]$ if necessary, we obtain in each case the desired witness of $\hat x \wedge a \vdash b$. 
\Prf

We denote the witness of $\hat x \wedge a \vdash b$ derived by pattern matching on $\phi: \bar \phi \to \hat \phi$ in the second half of the preceding proof by
$$\kappa_x (\phi): \hat x \wedge \bar \phi \to \hat \phi.$$
Now we pass from deductive systems to (ideal) categories. When we do so, it is necessary to ensure that the polynomial system over an indeterminate remains in our category. Hence we fix the following definition: 

\dfn\label{d.polyicat}
Let $\sC$ be an ideal cartesian closed category. %
Let $x$ be an indeterminate in $\sC$. 
To define the symbol
$$\sC(x),$$ 
observe that $\sC$ is equipped with the structure 
$$(s, t, \cdot, \vdash, \sI, \sV)$$
of a positive intuitionistic deductive system, when regarded as a generalized graph. % 
Take $\sK_{\sC(x)}$ to be the set of constant polynomials.\footnote{This definition restricts behavior of terminal arrows $\phi \vdash \psi$ for polynomials $\phi$ and $\psi$, but it will not make a difference for our purposes.} 
Now take the polynomial system $\sC[x]$ of Definition \ref{d.poly}, %
and then take the smallest equivalence relation $=_x$ of paths in $\sC[x]$ satisfying the conditions:
\begin{enumerate}
	\item If $f = g$ in $\sC$, then $f =_x g$ in $\sC(x)$,
	\item $(\phi \vdash \hat \psi ) \cdot \psi =_x (\phi \vdash \bar \phi)$ unless $\phi =_x \bar \psi$, in which case $(\phi \vdash \hat \psi) \cdot \psi = \psi$,
	\item $\psi \cdot (\bar \psi \vdash \phi) =_x (\hat \psi \vdash \phi) $ unless $\phi =_x \hat \psi$, in which case $\psi \cdot (\bar \psi \vdash \phi) =_x (\hat \psi \vdash \phi),$ unless $\psi \in \sK$ or $\bar \psi \vdash \phi \in \sK$, 
	\item For all $\phi, \psi \in \sC[x]$, if $(\chi \cdot \psi) \cdot \phi$ is defined, then 
		 $$(\chi \cdot \psi) \cdot \phi =_x \chi \cdot (\psi \cdot \phi),$$
	\item Composition $(\cdot)$, combination $\langle, \rangle$, and the turnstile $(\vdash)$ in $\sC(x)$ is substitutive in both arguments:
		\begin{enumerate}
			\item if $\phi =_x \psi$ then $\phi \vdash \chi =_x \psi \vdash \chi$ and $\chi \vdash \phi =_x \chi \vdash \psi$,
			\item if $\phi =_x \psi$ and $\phi \cdot \chi$ $\downarrow$ then $\phi \cdot \chi =_x \psi \cdot \chi$ and if $\chi' \cdot \phi$ $\downarrow$ then $\chi' \cdot \phi =_x \chi' \cdot \psi$,
			\item if $\phi =_x \psi$ and $\langle \phi, \chi \rangle$ $\downarrow$ then $\langle \phi, \chi \rangle =_x \langle \psi, \chi \rangle$ and $\langle \chi, \phi \rangle =_x \langle \chi, \psi \rangle$,
		\end{enumerate} 
	\item For all $\phi: a \to \trut$, $f =_x a \vdash \trut$,
	\item For all pairs $(a,b) \in \sC$ (viewed as a deductive system), if the unique good pair for $(a,b)$ is $(\pi_{a,b}, \pi'_{a,b})$ and any good evaluation $\epsilon_{a,b}$ is taken, then these are required to satisfy their usual equational properties in expressions involving $x$: 
		\begin{enumerate}
			\item if $\pi_{a,b} \langle \phi, \psi \rangle$ $\downarrow$ then $\pi_{a,b} \langle \phi, \psi \rangle =_x \phi$,
			\item if $\pi'_{a,b} \langle \phi, \psi \rangle$ $\downarrow$ then $\pi'_{a,b} \langle \phi, \psi \rangle =_x \psi$,
			\item if $\langle \pi_{a,b} \cdot \phi, \pi'_{a,b} \cdot \phi \rangle$ $\downarrow$ then $\langle \pi_{a,b} \cdot \phi, \pi'_{a,b} \cdot \phi \rangle =_x \phi$,
			\item if $\langle \phi \cdot \pi_{a,b} , \psi \cdot \pi'_{a,b} \rangle$ $\downarrow$ then $\langle \phi \cdot \pi_{a,b}, \psi \cdot \pi'_{a,b} \rangle = \phi \wedge \psi$,
			\item if $\epsilon \langle \phi^* \cdot \pi_{c,a}, \pi'_{c,a} \rangle$ $\downarrow$ then $\epsilon \langle \phi^* \cdot \pi_{c,a}, \pi'_{c,a} \rangle = \phi$, 
			\item if $(\epsilon \langle \phi \cdot \pi_{c,a}, \pi'_{c,a} \rangle )^*$ $\downarrow$ then $(\epsilon \langle \phi \cdot \pi_{c,a}, \pi'_{c,a} \rangle )^* = \phi$. %
		\end{enumerate}
\end{enumerate}
\Dfn

The construction of $\sC(x)$ is thus carried out closely following Lambek. 
By iterating the construction of Definition \ref{d.polyicat} we may define general polynomial systems $\sA[\vec{x}]$ and general polynomial categories $\sC(\vec{x})$. %
A {\em polynomial over $\sC$} is an element of $\sC(\vec x)$ for any sequence of indeterminates $\vec x$. %

The following properties are established in \cite{LaSc1} for ordinary cartesian closed categories. The proof in our setting is similar when source and target do not depend on $x$, but in general requires a recursive step:

\lem\label{l.fCompT}
Let $\sC$ be an ideal cartesian closed category. Then $\sC(x)$ is an ideal closed category, and moreover:
\begin{enumerate}
	\item For every ideal cartesian closed category $\sD$, for every $F: \sC \to \sD$, and for every $a: F(\bar x) \to F(\hat x)$ in $\sD$, %
		there exists a unique functor $\theta: \sC(x) \to \sD$ satisfying %
			$$ \theta(x) = a, \quad \theta(f) = F(f) \,\, \text{ for all } f \in \sC.$$
	\item As a consequence of (1), for every $a \in \sC$, there is a unique functor $S_x^a: \sC(x) \to \sC$ (called {\em substitution of $a$ for $x$}) satisfying
			$$S_x^a (x) = a, \qquad S_x^a (f) = f \,\, \text{ for all } f \in \sC.$$
\end{enumerate}
\Lem

\thm\label{t.fCompT}
Let $\sC$ be an ideal cartesian closed category, %
let $\phi \in \sC(x),$ where $\phi: \trut \to \hat \phi$. 
Then %
there exists a unique element $g:\hat x \to \hat{\phi}$ in $\sC$, such that
$$\phi = g \cdot x$$
in $\sC(x)$.
\Thm
\prf
The proof we give, following Lambek, proceeds by passing through $\sC[x]$, the polynomial generalized positive deductive system over $\sC$, and then verifying that one is able to mod out by $=_x$. 
First we show that $\kappa_x \phi$ has a new behavior because of $=_x$:

\lem\label{l.fCompT}
$\kappa_x \phi$ is a well-defined element of $\sC(x)$, satisfies
$$\kappa_x \phi \cdot \langle x , \trut \rangle = \phi,$$
and is the unique element of $\sC(x)$ that does so. 
\Lem
\prf 
One must check that
$$\eif \phi =_x \psi, \ethen \kappa_x \phi =_x \kappa_x \psi.$$
This requires checking each of the relations
We need only check the new case created by $\wedge$; the other cases can be checked as in \cite{LaSc1}. This follows from the definition of $\kappa_x$: for any $\phi, \psi$ in $\sC(x)$ we have $\phi \wedge \psi =_x \langle \phi \cdot \pi , \psi \cdot \pi' \rangle$. We verify that
\begin{align*}
\kappa_x (\phi \wedge \psi) 		&= \langle \kappa_x (\phi) \pi , \kappa_x (\psi) \pi' \rangle \\
						&= \langle \kappa_x (\phi \cdot \pi) , \kappa_x (\psi \cdot \pi') \rangle \\ %
						&= \kappa_x (\langle \phi \cdot \pi, \psi \cdot \pi' \rangle).     %
\end{align*}
from the definition of $\kappa_x$ for this case. 
The uniqueness of the choice of $\ksi(\phi)$ is the result of the following calculation in $\sC(x)$ \cite{LaSc1, LaK2}:
\begin{align*}
\kappa_x \phi 	&=_x \kappa_x (\tilde f \cdot \langle x , \trut \rangle) \\
			&=_x \tilde f \cdot \kappa_x ( \langle x , \trut ) \\
			&=_x \tilde f \cdot \langle \kappa_x x , \kappa_x \trut \rangle \\
			&=_x \tilde f \cdot \langle \pi_{\hat x, \trut}, \trut \cdot \pi'_{\hat x, \trut} \rangle \\
			&=_x \tilde f. \qedhere
\end{align*}
\Prf

Now we finish the proof of Theorem \ref{t.fCompT}. %
We define the element $g$ in $\sC(x)$ to be
$$g := \kappa_x \phi \cdot \beta,$$
where $\beta$ is just the obvious munging term, in fact $\beta \threeline \langle 1_{\hat x}, \hat{x} \vdash \trut \rangle$. %
Indeed, we have
\begin{align*}
g \cdot x 	
		&= \kappa_x \phi \cdot \langle 1_{\hat x}, \hat{x} \vdash \trut \rangle \\
		&= \kappa_x \phi \cdot \langle x, \trut \vdash \trut \rangle \\
		&= \kappa_x \phi \cdot \langle x, \trut \rangle \\
		&= \phi
\end{align*}
by Lemma \ref{l.fCompT}. %
For uniqueness of $g$, suppose that $\tilde g \in \sC$ satisfies $\tilde g \cdot x = \phi$ in $\sC(x)$. We calculate
\begin{align*}
\kappa_x (\phi) \cdot \beta 	
		&= \kappa_x (\tilde g \cdot x) \cdot \beta \\
		&= \kappa_x (\tilde g \cdot x) \cdot \langle 1_{\hat x}, \hat x \vdash \trut \rangle \\
		&= \kappa_x (\tilde g) \cdot \langle \pi_{\hat x, \trut}, \kappa_x x \rangle \cdot \langle 1_{\hat x} , \hat x \vdash \trut \rangle \\
		&= \tilde g \cdot \pi'_{\hat x, \hat x} \langle \pi_{\hat x, \trut} , \pi_{\hat x, \trut} \rangle \cdot \langle 1_{\hat x} , \hat x \vdash \trut \rangle \\
		&= \tilde g \cdot \pi_{\hat x, \trut} \cdot \langle 1_{\hat x}, \hat x \vdash \trut \rangle \\
		&= \tilde g \cdot 1_{\hat x} \\
		&= \tilde g.
\end{align*}
But $\kappa_x (\phi) \cdot \beta = g$ by definition of $g$. So $g = \tilde g$, and $g$ is unique. 
\Prf

From Theorem \ref{t.fCompT} we define notation (to resemble a counit) $\varepsilon_x \phi: \hat x \to \hat \phi$ by
$$\varepsilon_x \phi := g = \kappa_x (\phi) \cdot \beta.$$
Theorem \ref{t.fCompT} has the following corollary:

\cor\label{c.fCompT}
Let $\sC$ be an ideal cartesian closed category, %
and let $\phi \in \sC(x)$ have source $\trut$. %
Then there exists a unique element $h: \trut \to (\hat x \vdash \hat \phi)$ such that 
$$\phi =_x \epsilon \cdot \langle h, x \rangle$$
in $\sC(x)$. %
\Cor
\prf
This is obtained by taking the name of the element $g$ of Theorem \ref{t.fCompT}: that is, take
\[
h = \name{g}. \qedhere
\]
\Prf

From Corollary \ref{c.fCompT} we define notation $\lambda_x \phi : \trut \to (\hat x \vdash \hat \phi)$ by
\[
\lambda_x \phi := h = \name{\kappa_x (\phi) \cdot \langle 1_{\hat x} , \hat x \vdash \trut \rangle} 
\]

As an aside, we observe from the proofs of Theorem \ref{t.dedt} and \ref{t.fCompT} that $\wedge$'s identity in categories suggests whether the symbol may be sugared out of generalized deduction systems entirely. This would mean $\langle, \rangle$ would be defined as a basic operation subject to an equational axiom: %
$$t(\langle a, b \rangle) = \langle \hat a \pi, \hat b \pi' \rangle.$$
In this case $\pi$ and $\pi'$ must satisfy a self-referential axiom:
$$s(\pi) = s(\pi') = \langle a \cdot \pi, b \cdot \pi' \rangle.$$

%

%

%

% lambda calculus
\section{Typed Lambda Calculus and the Main Correspondence}\label{s.lam}

In this section we will finally observe what happens on the syntactic side of the correspondence after generalizing semantics. As it turns out,  types acquire a richer structure and simultaneously assume the role of function constants. %
By a {\em generalized lambda calculus} (Definition \ref{d.lam}) we refer to the simplest such type system possible: %
we do not make mention of natural numbers objects (see \cite{LaSc1}), Boolean types, or other features that may appear in applications of lambda calculus. %
The next definition is not used in the sequel. It is included in order to establish a basis for defining variables before making Definition \ref{d.lam}.

\dfn\label{d.prelam}
A {\em pre-generalized typed lambda calculus} is a structure
$$(\Lambda, \sT_\Lambda, \sS_\Lambda, s,t, \cdot, \vdash, %
\trut, \wedge, \ty, \name{}, *, ()^\cdot, \pi, \pi', \wr, \langle,\rangle, \lambda, \sV_\Lambda)$$
where
\begin{enumerate}
	\item $\Lambda$ is a set,
	\item $\sT_\Lambda$ and $\sS_\Lambda$ are disjoint subsets of $\Lambda$ and $\sT_\Lambda \cup \sS_\Lambda = \Lambda$,
	\item $\sV_\Lambda$ is a subset of $\Lambda$,	
	\item the system
		$$(\sT_\Lambda, s,t,\cdot,\vdash, \sV')$$
		is an ideal category, where $\sV' = \sV_\Lambda \cap \sT_\Lambda$,
	\item $\trut$ is a designated element of $\sT_\Lambda$,
	\item $\wedge$ is a mapping $\sT_\Lambda \times \sT_\Lambda \to \sT_\Lambda$,
	\item $\name{}$ is a mapping $\sT_\Lambda \to \sS_\Lambda$,
	\item $\ty$ is a mapping $\sS_\Lambda \to \sT_\Lambda$,
	\item and in $\sS_\Lambda$:
	\begin{enumerate}
		\item $*$ is a designated element of $\sS_\Lambda$,
		\item $()^\cdot$ is a mapping $\Lambda \to \sS_\Lambda$,
		\item $\pi$, $\pi'$ are partially defined mappings $\sS_\Lambda \to \sS_\Lambda$,
		\item $\wr$ and $\langle,\rangle$ are partially defined mappings $\sS_\Lambda \times \sS_\Lambda \to \sS_\Lambda$,
		\item $\lambda$ is a mapping $\sX \times \sS \to \sS$, where $\sX$ is defined below,
	\end{enumerate}
\end{enumerate}
subject to the conditions
\begin{enumerate}
	\item % trut
		$\hat \trut = \bar \trut = \trut,$
 	\item % pi, pi'
		for all $s \in \sS_\Lambda$, $\pi(s)$ $\downarrow$ iff $\pi'(s)$ $\downarrow$ iff there exist $A,B \in \sT_\Lambda$ such that $\ty(s) = A \wedge B$, 
	\item % wr
		$s \wr t$ $\downarrow$ iff there exist $A,B \in \sT_\Lambda$ such that $\ty(s) = A \vdash B$ and $\ty(t) = A$,
	\item % langle,rangle
		$\langle s,t \rangle$ $\downarrow$ iff $\ty(s) = \ty(t)$,
\end{enumerate}
typing conditions
\begin{enumerate}
	\item % name
		$\ty(\name{A}) = \bar A \vdash \hat A$,
 	\item % *
		$\ty(*) = \trut$,
 	\item % ^cdot
		for all $\alpha \in \Lambda$, $\ty(\alpha^\cdot) = \ty(\alpha),$
 	\item % pi, pi'
		if $s \in \sS_\Lambda$ and $\ty(s) = A \wedge B$, then $\ty(\pi(s)) = A$ and $\ty(\pi'(s)) = B$,
	\item % wr
		if $s \wr t$ $\downarrow$, then $\ty(s \wr t) = \widehat{\ty(s)}$,
	\item % langle,rangle
		if $\langle s, t \rangle$ $\downarrow$, then $\ty(\langle s, t \rangle) = \ty(s) \wedge \ty(t)$,
	\item % \lambda 
		if $\lambda(x, s)$ $\downarrow$, then $\ty(\lambda(x, s)) = \ty(x) \vdash \ty(s)$,
\end{enumerate}
and the validities
\begin{enumerate}
	\item % trut
		$* \in \sV_\Lambda$,
	\item % wedge
		if $A,B \in \sV_\Lambda$, then $A \wedge B \in \sV_\Lambda$,
	\item % ty
		(witnesses, propositions-as-types) if $s \in \sV_\Lambda$, then $\ty(s) \in \sV_\Lambda$,
	\item % name
		If $A \in \sV_\Lambda$, then $\name{A} \in \sV_\Lambda$.
 	\item % *
		$* \in \sV_\Lambda$,
 %	\item % ^cdot		
 	\item % pi, pi'
		if $c \in \sV_\Lambda$ and $\pi(c), \pi'(c)$ $\downarrow$, then $\pi(c), \pi'(c) \in \sV_\Lambda$,
	\item % wr
		if $a,f \in \sV$, then $f \wr a \in \sV$,
	\item % langle,rangle
		$a,b \in \sV_\Lambda$ implies $\langle a, b \rangle \in \sV_\Lambda$,
	\item % \lambda 
		if $s \in \sV_\Lambda$, then $\lambda(x,s) \in \sV_\Lambda$. %
\end{enumerate}
\Dfn

Note that many type theories, e.g. \cite{MoI1}, include {\em function constants} $f: A \to B$ as well as terms and types; in this formalism (guided by the new semantics) function constants are indistinguishable from types, and together with objects they form a category. Types behave as function constants via the derived operation
$$A \star s := \name{A} \wr s.$$
The operation $\name{}$ is used not only here but also in the construction of $\boC \Lambda$ in Definition \ref{d.catcon}. 

Elements of $\sT_\Lambda$ are called {\em types}, and elements of $\sS_\Lambda$ are called {\em terms}. For a term $s$, the element $\ty(s)$ of $\sT_\Lambda$ is called the {\em type of $s$.} %
We may write $s:T$ to denote the relation $\ty(s) = T$.
A term of the form $\alpha^\cdot$ for some $\alpha$ (which may be a type or a term) is called a {\em variable}. %
We may iterate the operation $()^\cdot$, %
and we do not allow $()^\cdot$ to be substitutive in its argument. %
Therefore we may assume that the symbol $x_i$ %
unpacks to %
$((\dots ((A)^\cdot )^\cdot \dots )^\cdot )^\cdot$. %
In this way, we have a countable stock $x_1, x_2, \dots$ of distinct ``standard'' variables for each type $A$. %
For technical reasons (see below, before Definition \ref{d.lam}), we take these standard variables to be the only variables of $\Lambda$, 
and we place the obvious (total) ordering on variables of each type. %
A variable $x_i$ is {\em free} in a term if it appears in the term, unless it appears but only within a well-formed expression of the form $\lambda(x_i, s)$. In this case we say it appears {\em captured} or {\em bound}. 
We define the mapping on terms
$$\FV(s) = \set{x \in \sX \mid x \text{ appears free in $s$, and } x \nin \sV_\Lambda},$$
where the phrase ``appears free'' has its usual meaning, except that we assume that no variable {\em appears free} in any type. So, for example, for all types $A$, $\FV(\name{A})$ is empty. %
If $s$ is a term, $x$ is a variable, and $t$ is a term whose type is the same as the type of $x$ %
we define notation
$$s[x/t]$$
to be the term $s$ with the variable $x$ replaced by $t'$ in each instance where it does not appear bound in $s$, %
where $t'$ is $t$ with any variable $y \in \FV(t)$ that appears captured in $s$, that is, 
$$y \in \FV(t) \cap \CAP(s),$$
where $\CAP(s)$ is the set of variables appearing captured in $s$, %
replaced by a variable of the same type that is not in the set $\VAR(s) \cup \VAR(t)$ of variables appearing in either $s$ or $t$. %
These choices are made in {\em the simplest order-preserving way}, by which is meant that once the set of variables to be changed is found, the entire set is incremented by the smallest positive integer such that the set of variables so generated is not in $\VAR(s) \cup \VAR(t)$. %
These incrementing operations are associative, as is the substitution operation itself. Hence we have
$$s[x/t][y/r] = s[x/t[y/r]]$$
for all terms $s,r,t$ and variables $x,y$. %
We may often ignore the extra step involving $t'$, for it is only necessary because we have not set terms $s$ and $s'$ equal in $\Lambda$ which are the same up to one or more free variables (a form of $\alpha$-conversion) in $\prelambdaCalc$ or in the category $\lambdaCalc$ defined next. %
Note that a morphism in $\prelambdaCalc$ sends closed terms to closed terms. %

\dfn\label{d.lam}
A {\em generalized typed lambda calculus} is %
a pre-generalized typed lambda calculus %
on which there is an equality relation on the terms $\sS_\Lambda$ of $\Lambda$ defined as follows: %
Let $\sP$ be the finite power set $\sP_{fi} (\sX)$ of $\sX$. %
For each finite set $\bar x = \set{x_1, \dots, x_n}$ in $\sP$, %
let 
$$\sR(\Lambda, \bar x) := \set{s \in \sS_\Lambda \mid \FV(s) \lies \bar x}.$$
We define the relation $=_{\bar x}$ on $\sR(\Lambda, \bar x)$ 
to be the smallest equivalence relation that satisfies
\begin{enumerate}
	\item $=_{\bar x}$ is reflexive, symmetric, and transitive,
	\item Substitutivity conditions:
	\begin{enumerate}
		\item % pi,pi'
			if $s =_{\bar x} t$, and $\pi(s)$ $\downarrow$, then $\pi(s) =_{\bar x} \pi(t)$, and $\pi'(s) =_{\bar x} \pi'(t)$,
		\item % wr 
			if $s =_{\bar x} t$, then $s \wr r =_{\bar x} t \wr r$ and $u \wr s =_{\bar x} u \wr t$ whenever these expressions are well-defined, 
		\item % langle
			if $s =_{\bar x} t$, and $\langle s,r \rangle$ $\downarrow$, then $\langle s,r \rangle =_{\bar x} \langle t,r \rangle$, and similarly in the second argument,
		\item % lambda
			if $s =_{\bar x} t$, then $s \wr r =_{\bar x} t \wr r$ and $u \wr s =_{\bar x} u \wr t$ whenever these expressions are well-defined, 
	\end{enumerate}
	\item for all $s:\trut$, $s =_{\bar x} *$,
	\item for all $a : A, b: B,$ 
		$$\pi(\langle a, b \rangle) =_{\bar x} a,$$
		$$\pi'(\langle a, b \rangle) =_{\bar x} b,$$
	\item for all $c: A \wedge B$,
		$$\langle \pi(c), \pi'(c) \rangle =_{\bar x} c,$$
	\item For all terms $s \in \sS_\Lambda$, terms $a \in \sS_\Lambda$, and variable $x$ that may appear in $\bar x$,
	\begin{enumerate}
		\item $(\lambda(x,s)) \wr a =_{\bar x} s[x/a],$
		\item $\lambda(x, s \wr x) =_{\bar x} s$,
		\item if $\FV(s) = \set{x}$, there exists a unique $A \in \sT_\Lambda$ such that $s =_{\set{x}} \name{A} \wr x$. \label{ax.unname}
		\item ($\alpha$-conversion for lambda terms) 
		$$\lambda(y, s) =_{\bar x} \lambda(y', s[y/y'])$$
		if $\ty(y) = \ty(y')$ and $y' \nin \FV(s)$.
	\end{enumerate}
\end{enumerate}
We observe that $\FV()$ is still well-defined. %
We denote by 
$$s^\natural$$ 
the type $A$ given by Axiom \ref{ax.unname}. % 
We impose the condition on the $=_{\bar x}$'s that: 
\begin{enumerate}
	\item if $\bar x \lies \bar y$ then for all $s,t \in \sR(\Lambda, \bar y)$, $s =_{\bar x} t$ implies $s =_{\bar y} t$.
\end{enumerate}
Because %
(1) $s =_{\FV s} s$, and %
(2) if $s =_{\bar x} t$ and $s =_{\bar y} t$, then there exists a finite set $\bar z$ such that $s =_{\bar z} t$ and $\bar x, \bar y \lies \bar z$, %
we may define an equivalence relation {\em equality in $\sS_\Lambda$} on the set $\sS_\Lambda$ of terms of $\Lambda$ by
$$s = t \,\,\,\text{ if }\,\, s =_{\bar x} t \text{ for some $\bar x$ in $\sP$}.$$
A {\em morphism} $\Phi: \Lambda \to \Mu$ of generalized typed lambda calculi, also called a {\em translation}, is a mapping 
$$\Phi: \Lambda \to \Mu$$
that satisfies the following, where equalities between terms are interpreted as equality in $\sS_\Lambda$:
\begin{enumerate}
	\item for all $A \in \sT_\Lambda, s \in \sS_\Lambda$, $\Phi(A) \in \sT_\Mu$ and $\Phi(s) \in \sS_\Mu,$
	\item the restriction of $\Phi$ to $\sT_\Lambda$ is a morphism of ideal categories that satisfies
		$$\Phi(\trut_\Lambda) = \trut_\Mu,$$
		$$\Phi(A \wedge B) = \Phi(A) \wedge \Phi(B),$$
	\item if $s =_{\bar x} t$, then $\Phi(s) =_{\Phi(\bar x)} \Phi(t)$. \label{ax.preservesequalities}
	\item % ty
		$\Phi(\ty(s)) = \ty(\Phi(s)),$
	\item % name
		$\Phi(\name{A}) = \name{\Phi(A)},$
	\item % *
		$\Phi(*) = *$, 
	\item % ^cdot
		for all $\alpha \in \Lambda$, $\Phi(\alpha^\cdot) = \Phi(\alpha)^\cdot,$
	\item % pi, pi'
		$\Phi(\pi(c)) = \pi(\Phi(c),$ and $\Phi(\pi'(c)) = \pi'(\Phi(c)),$
	\item % wr
		$\Phi(s \wr t) = \Phi(s) \wr \Phi(t),$
	\item % langle,rangle
		$\Phi(\langle s, t \rangle) = \langle \Phi(s), \Phi(t) \rangle,$
	\item % lambda
		$\Phi(\lambda(x,s)) = \lambda(\Phi(x), \Phi(s)).$
\end{enumerate}
As a consequence of (\ref{ax.preservesequalities}), $\Phi$ preserves equalities in $\Lambda$:
$$s = t \eimplies \Phi(s) = \Phi(t).$$
This gives a category $\lambdaCalc$ of generalized typed lambda calculi.
\Dfn

Given a generalized typed lambda calculus $\Lambda$, we can construct an ideal cartesian closed category using Theorem \ref{t.fCompT}:

\dfn\label{d.catcon}
Let $\Lambda$ be a typed lambda calculus. %
Let $\sB_\Lambda$ be the set of {\em bulletins} in $\Lambda$, that is, the set of terms in $\Lambda$ that have only one free variable. %
Also for $A \in \sT_\Lambda$, let 
$$\settl{\bullet}{A} := \settl{x}{\name{A}\wr x}, \quad x : \bar A,$$
that is, a symbol $\settl{x}{s}$ where $x$ is a variable of type $\bar A$, and $s$ is the term $\name{A} \wr x$. %
By Axiom (\ref{ax.unname}) of Definition \ref{d.lam} we may identify these symbols with types in $\Lambda$. 
We define $\boC\Lambda$ to be the set
$$\boC\Lambda := \set{\settl{x}{s} \mid s \in \sB_\Lambda, \text{ $x$ a variable}},$$ 
of symbols $\settl{x}{s}$ for variable $x$ and bulletin $s$, equipped with the structure
\begin{align*}
		\overline{\settl{x}{s}} &:= \settl{\bullet}{\ty(x)}, \\ %		
		\widehat{\settl{x}{s}} &:= \settl{\bullet}{\ty(s)}, \\ %
		\settl{x}{s} \cdot \settl{y}{t} &:= \settl{y'}{s[x/t]}, \\
		\settl{x}{s} \vdash \settl{y}{t} &:= \settl{u}{v}, \quad u:\ty(s), v:\ty(t), 
\end{align*}
where
$$y' = 
	\begin{cases} 	
				y 
						& \text{ if $\FV(t)$ is empty}, \\ 
				inc_n(y)	
						& \text{ if $\FV(t) = \set{u}$ and $\FV(s[x/t]) = \set{inc_n (u)}$,}
	\end{cases}
$$
where $inc_n$ is the modification of the variable described after Definition \ref{d.prelam}. %
Let $\sK_{\boC\Lambda}$ be the set of symbols $\settl{x}{k}$ where $k$ is a constant in $\Lambda$, that is, $\FV(k) = \nll$. %
Let equality of symbols in $\boC\Lambda$ be defined by
\begin{enumerate}
	\item $\settl{x}{s} = \settl{y}{t}$ if $\ty(x) = \ty(y), \ty(s) = \ty(t),$ and there is $z:\ty(x)$ such that $s[x/z] = t$, 
	\item $\settl{x}{s} = \settl{x}{u}, \quad u:\ty{s}, \,\,$ \text{ if $\FV(s) = \set{y}$ and $y \neq x$}, 
	\item if $\ty(s) = \trut$, then $\settl{x}{s} = \settl{x}{*}$, 
	\item for all bulletins $s$ and all variables $x,y$ of the same type, $\settl{x}{s} = \settl{y}{s[x/y]}$. 
\end{enumerate}
This gives an ideal category $\boC\Lambda$, where %
the identity of $\settl{x}{s}$ is 
$$1_{\settl{x}{s}} = \settl{y}{y}, \quad y:s^{\natural},$$
terminal arrows are of the form 
$$\settl{y}{*},$$
and types (in the sense of section \ref{s.th}) are of the form
$$\settl{x}{y}, \quad x \neq y.$$
Validities defining $\boC\Lambda$ are the evident ones based on Definition \ref{d.icat}. 
\Dfn

We have an ideal category $\boC\Lambda$, but we have not directly made any assumptions about the category $\sT_\Lambda$. Nevertheless, we have:

\prop\label{p.catcon}
$\boC\Lambda$ is an ideal cartesian closed category. 
\Prop
\prf
Set
\begin{align*}
		\trut &:= \settl{u}{*}, \quad u:\trut_\Lambda,\\
		\settl{x}{s} \wedge \settl{y}{t} &:= \settl{z}{\langle s \wr \pi(z), t \wr \pi'(z) \rangle}, \\
		\langle \settl{x}{s}, \settl{y}{t} \rangle &:= \settl{z}{\langle s[x/z], t[y/z] \rangle}, \\
		\settl{z}{s}^* &:= \settl{x}{\lambda(y,s \wr \langle x,y \rangle) }, \quad \text{ where $z:A \times B, x:A$}, \\
		\pi &:= \settl{z}{\pi(z)}, \\
		\pi' &:= \settl{z}{\pi'(z)},\\
		\epsilon &:= \settl{z}{\pi(z) \wr \pi'(z)}, 
\end{align*}
with validities as needed (Definition \ref{d.iccc}). 
\Prf

We can also construct a typed lambda calculus from the data of a cartesian closed category:

\dfn\label{d.internallanguage}
Let $\sC$ be an ideal cartesian closed category. We define the symbol $\boL\sC$ as follows:
\begin{enumerate}
	\item The set of types of $\boL\sC$ is the set of symbols $A_f$ indexed by elements $f \in \sC$:
		$$\sT_{\boL\sC} := \set{A_f \mid f \in \sC},$$
		in fact we set $A_f = f$ and take $\sC$ itself as the set of types (this is needed for the proof of Theorem \ref{t.chl}), however, we use the notation $A_f$ at times when it seems to lessen the potential for confusion.
	\item The set of terms of $\boL\sC$ is the set of polynomials $\phi$ over $\sC$ sourced at $\trut$, that is, %
		$$\sS_{\boL\sC} := \set{\phi \mid \phi \in \sC[\vec x] \text{ for some $\vec x$, } \hat{\phi} \text{ is in $\sC$, and } \bar{\phi} = \trut}, $$
		where we assume %
		that indeterminates have internal structure given by the syntax $()^\cdot$. 
	\item Define
		\begin{align*}
			\ty(\phi) &:= A_{\hat \phi}, \\
		 	s(A_f) &:= A_{sf}, \\
			t(A_f) &:= A_{tf}, \\
			A_f \cdot A_g &:= A_{f \cdot g}, \\
			A_f \wedge A_g &:= A_{f \wedge g}, \\
			A_f \vdash A_g &:= A_{f \vdash g}, \\
			\name{A_f} &:= \name{f}, \quad \text{ the name of $f$,} \\
			\trut_{\boL\sC} &:= \trut_\sC \vdash \trut_\sC, \\
			* &:= \trut_\sC, \\
			\sK_{\boL\sC} &\text{ is the set of constant polynomials.} %
		\end{align*}
	\item if $\phi$ is a bulletin in $x$ over $\sC$, then let 
		$$\phi^\natural := A_{\epsilon_x \phi}.$$
	\item $\sV_{\boL\sC}$ is the set $\set{A_f  \mid f \in \sV_\sC}$ %
	joined with the set of valid constant terms, %
	joined with the set of polynomials valid according to Definition \ref{d.poly}. %
	\item Define
		\begin{align*}	
			\trut_{\boL\sC} 	&:= \trut \vdash \trut, \\
			*_{\boL\sC} 	&:= \trut,
		\end{align*}
\end{enumerate}
Then we have a pre-generalized typed lambda calculus. %
We make from this a generalized typed lambda calculus by imposing the equality relation on terms inherited from equality in $\sC(\vec x)$:
the equality relation $=_{\vec x}$ is defined to be equality in $\sC(\vec x)$, along with the usual inclusions of polynomial systems in one another. 
\Dfn

$\boL\sC$ is called the {\em internal language} of the ideal cartesian closed category $\sC$. %
Next, we verify that these constructions are functorial: 

\prop\label{p.functor}
We have the following:
\begin{enumerate}
	\item $\boC$ is a functor from $\lambdaCalc$ to ${\bf ICCC}$. % 
	\item $\boL$ is a functor from ${\bf ICCC}$ to $\lambdaCalc$. %
\end{enumerate}
\Prop
\prf
Given $\Phi: \Lambda \to \Lambda'$, we define $\boC\Phi : \boC\Lambda \to \boC\Lambda'$ by
$$\boC\Phi \settl{x}{s} := \settl{\Phi(x)}{\Phi(s)}$$
for %
$\settl{x}{s} \in \boC\Lambda.$ 
Now we check that $\boC \Phi$ is a morphism in ${\bf ICCC}$, %
and that %
$\boC$ is a functor (Definition \ref{d.functor}). %

Let $F: \sC \to \sD$ in ${\bf ICCC}$. Define a mapping $\boL F: \boL(\sC) \to \boL(\sD)$ by
\begin{align*}
\boL F(A_f) &:= A_{F(f)}, \\
\boL F ( \alpha^\cdot) &:= (\boL F(\alpha))^\cdot, 
\end{align*}
and extend $F$ from $\sC$ to polynomials over $\sC$ in the most straightforward way. % 
Now we check that $\boL F$ is a morphism in $\lambdaCalc$, and that $\boL$ is indeed a functor. %
\Prf

\dfn\label{d.eta}
Let $\Lambda$ be a generalized typed lambda calculus. Define a mapping $\Lambda$ to $\boL\boC\Lambda$ %
by defining, in the pre-generalized typed lambda calculus $\Lambda_0$ obtained by ignoring equalities in $\sS_\Lambda$, 
\begin{align*}
\eta_\Lambda(A) &:= A_{\settl{\bullet}{A}}	& A \in \sT_\Lambda, % 
		 			\\
\eta_\Lambda(k) &:= \settl{x}{k}, 			& k \in \sS_\Lambda, \FV(k) = \nll, \ty(x) = \trut_{\boL\boC\Lambda},	\\
\eta_\Lambda(\alpha^\cdot) &:= (\eta_\Lambda(\alpha))^\cdot,					& \alpha \in \Lambda,		\\
\eta_\Lambda(\pi(\phi)) &:= \pi(\eta_\Lambda \phi ), 												& \\
\eta_\Lambda(\pi'(\phi)) &:= \pi'(\eta_\Lambda \phi ),												& \\
\eta_\Lambda(\langle \phi, \psi \rangle) &:= \langle \eta_\Lambda (\phi) , \eta_\Lambda( \psi) \rangle 			& \\
\eta_\Lambda(\phi \wr \psi) &:= \eta_\Lambda (\phi) \wr \eta_\Lambda (\psi)								& \\
\eta_\Lambda(\lambda(x, \phi)) &:= \lambda(\eta_\Lambda(x), \eta_\Lambda (\phi))			& x \in \sX_\Lambda 
\end{align*}
The map $\eta_\Lambda$ is well-defined upon passage to $\Lambda$, since analogous equalities between polynomials hold in both $\Lambda$ and $\boL\boC\Lambda$. 
\Dfn

An alternative approach (really the same) to Definition \ref{d.eta} is via an isomorphism with a lambda calculus with parameter \cite{LaSc1}: 

\dfn\label{d.indeterminatevariable}
Let $\Lambda$ be a generalized typed lambda calculus, and let $x \in \sX_\Lambda$ be a variable. We define the symbol 
$$\Lambda_x$$
to be the generalized typed lambda calculus is defined exactly as $\Lambda$, except that
$$\sV_{\Lambda_x} := \set{x} \cup \sV,$$
that is, $x$ is taken to be a validating term in $\Lambda_x$. 
\Dfn

Intuitively, $\Lambda_x$ is $\Lambda$ with $x$ treated as a constant instead of as a variable. %

\lem\label{l.indeterminatevariable}
Let $\sC$ be an ideal cartesian closed category, and let $x$ be an indeterminate (with the variable syntax). Then the polynomial category 
$\boC\Lambda(x)$ over $\boC\Lambda$ is isomorphic to $\boC\Lambda_x$ in ${\bf ICCC}$. %
\Lem
\prf
By Proposition \ref{l.fCompT}, we need only check that $\boC\Lambda_{x}$ has the desired universal property of $\boC\Lambda(x)$. %
See \cite{LaSc1}. %
\Prf

Using Lemma \ref{l.indeterminatevariable}, we can identify polynomials $\tilde \phi$ over $\boC\Lambda$ with the corresponding symbol $\settl{u:\trut}{\phi(\vec x)}$ in $\boC\Lambda_{\vec x}$, where $\vec x = \FV(\phi)$ corresponds to the free variables $\ksi_1, \dots, \ksi_n$ of $\tilde \phi$ over $\boC\Lambda$ via the isomorphism. 

Finally, we have an extension of Lambek's equivalence between simply typed lambda calculi and cartesian closed categories:

\thm\label{t.chl}
The functors $\boC$ and $\boL$ form an equivalence
\[
\raisebox{-0.5\height}{\includegraphics{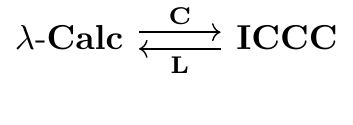}}
\]

\vspace{-15pt}\noindent
between $\lambdaCalc$ and ${\bf ICCC}$. 
\Thm
\prf
For $\sD$ in ${\bf ICCC}$, define $\varepsilon_\sD: \boC\boL \sD \to \sD$ to be the map 
$$\varepsilon_\sD : \settl{x}{\phi} \mapsto 
		\begin{cases} 
			\varepsilon_x \phi, & \text{if $\FV(\phi) = \set{x},$ or $\FV(\phi)$ is empty, or $\ty(\phi) = \trut_\Lambda$,} \\
			\hat x \vdash \hat \phi \,\,\text{ in $\sC$, } & \text{otherwise.} 
		\end{cases}
$$
This map is well-defined since if $\settl{x}{\phi} = \settl{y}{\psi}$, then $\phi[x/z] = \psi[y/z]$, where $z$ does not appear in $\phi$ or $\psi$. Let these be $\phi(z), \psi(z)$. Then $\varepsilon_z \phi(z) = \varepsilon_z \psi(z)$. But $z$ is eliminated by evaluation, so $\varepsilon_x \phi = \varepsilon_z \phi(z) = \varepsilon_z \psi(z) = \varepsilon_x \psi.$
Let $F: \sC \to \sD$ in ${\bf ICCC}$. Then to check that $\varepsilon: \sD \mapsto \varepsilon_\sD$ is a natural transformation, that is,
$$\varepsilon(\sD) \of \boC\boL(F) = F \of \varepsilon(\sC),$$
we check that for every $\settl{x}{\phi}$ in $\boC\boL \sC$, where $\phi$ is a bulletin in $x$ over $\sC$,
$$\varepsilon_\sD (\boC\boL F(\settl{x}{\phi})) = F(\varepsilon_\sC (\settl{x}{\phi})).$$
If $\phi$ is a non-constant bulletin in a variable different than the variable appearing in the symbol, then
\begin{align*}
\varepsilon_\sD (\boC\boL F \settl{x}{\phi}) 
	&= \varepsilon_\sD (\boC\boL F \settl{x}{y} ) \\
	&= \varepsilon_\sD \settl{x' : F(\ty(x))}{y' : F(\ty(y))} \\
	&= F(\ty(x)) \vdash F(\ty(y)) \\
	&= F( \ty(x) \vdash \ty(y) ) \\
	&= F(\varepsilon_\sC \settl{x}{\phi}).
\end{align*}
In the other cases, %
this reduces to checking that %
$$F(\varepsilon_x \phi) = \varepsilon_z \boL F \phi,$$
where $\boL F(x) \threeline z.$ We proceed by cases as in the proof of Theorem \ref{t.dedt}: % 
if $\phi$ is a constant $k: \trut \to \hat k$, then 
\begin{align*}
F \varepsilon_x \phi 
	&= F(k) \cdot F(\pi'_{\trut, \trut} \cdot \langle 1_\trut , \trut \vdash \trut \rangle ) \\
	&= F(k) \cdot 1_\trut \\
	&= F(k) \\
	&= \boL F(k) \\
	&= \epsilon_z \boL F(k). 
\end{align*}
If $\phi$ is a variable $x: \trut \to \hat x$ equal to the variable captured by the symbol, then
\begin{align*}
F \varepsilon_x \phi
	&= F \varepsilon_x x \\
	&= F( \pi_{\trut, \hat x} \cdot \langle 1_\trut, \hat x \vdash \trut \rangle ) \\
	&= \pi_{\trut, \widehat{F(x)}} \cdot \langle 1_\trut, \widehat{F(x)} \vdash \trut \rangle ) \\
	&= \varepsilon_{\boL F (x)} \,\boL F(x). 
\end{align*}
The other cases are similar. 

For a generalized typed lambda calculus $\Lambda$ in $\lambdaCalc$, define $\eta(\Lambda) := \eta_\Lambda$ of Definition \ref{d.eta}.
To show that $\eta$ is a natural transformation, let $\Phi: \Lambda \to \Mu$ in $\lambdaCalc$. Then
$$\eta(\Mu) \of \Phi = \boL \boC(\Phi) \of \eta(\Lambda)$$
becomes, for types, 
$$\eta_\Mu (\Phi(A)) = \boL\boC\Phi (\eta_\Lambda (A)),$$
which is easily verified. Indeed,
\begin{align*}
\boL\boC\Phi(\eta_\Lambda (A))
	&= \boL\boC\Phi(A_{\settl{\bullet}{A}}) \\
	&= A_{\boC\Phi{\settl{\bullet}{A}}} \\
	&= A_{\settl{z}{\Phi(\name{A}) \wr z}}, \quad \ty z = \Phi(\bar A) = \overline{\Phi(A)}, \\
	&= A_{\settl{z}{\name{\Phi(A)} \wr z}} \\
	&= \eta_\Mu (\Phi(A)).
\end{align*}
For terms, we proceed by induction on the length of a term $s$ of $\Lambda$. %
If $s = k$ is a constant term (of length zero), 
\begin{align*}
\eta_\Mu (\Phi(k)) 
	&= \settl{u}{ \Phi k } \quad u:\trut \\
	&= \settl{u }{ \Phi k } \quad u: \Phi(\trut) \text{ since $\Phi(\trut) = \trut$} \\
	&= \boC \Phi \settl{x}{k} \\
	&= \boL \boC \Phi \settl{x}{k} \\
	&= \boL\boC\Phi (\eta_\Lambda(k)).
\end{align*}
If $s = x$, a variable of type $A$, then
\begin{align*}
\boL\boC\Phi(\eta_\Lambda(x)). 
	&= \boL \boC \Phi(\ksi), \quad \ksi: \settl{\bullet}{A} \\
	&= \Phi\ksi, \quad \Phi\ksi : \settl{\bullet}{\Phi A} \\
	&= \eta_\Mu \Phi (x). 
\end{align*}
We can similarly check the other cases $\pi(t), \pi'(t), t \wr r, \langle t, r \rangle, \lambda(y, t)$. 

Both $\eta_\Lambda$ and $\varepsilon_\Lambda$ are invertible as maps. %
Indeed, by Theorem \ref{t.fCompT}, $\varepsilon$ is injective, and also surjective (since $g \cdot y$ is itself a polynomial). %
To show that $\eta_\Lambda$ is invertible, we use Lemma \ref{l.indeterminatevariable}: if $\phi$ is a polynomial over $\boC\Lambda$ in variables $x_1, \dots, x_n$, we pass via the isomorphism of Lemma \ref{l.indeterminatevariable} from $\phi$ to an element $\phi'$ in $\boC\Lambda_{x_1, \dots, x_n}$ of the form $\settl{y}{t}$. Now note that $\eta_\Lambda(t) = \phi$, so $\eta_\Lambda$ is surjective. On the other hand if $\eta_\Lambda s = \eta_\Lambda t$, for two terms $s,t \in \sS_\Lambda$, then 
$\settl{u:\trut}{s} = \settl{u:\trut}{t} \quad \text{ in $\boC\Lambda_{x_1, \dots, x_n}$.} $ %
so $s = t$ as terms over $\Lambda$, by definition of equality in $\sS_{\boC\Lambda_{x_1, \dots, x_n}}$. %

Next we check (cf. Definition \ref{d.adjunction}) that the triangle laws hold. 
Let $\sC$ be in ${\bf ICCC}$. For a type $A_f$ in $\boL\sC$, 
\begin{align*}
\boL\varepsilon_\sC (\eta_{\boL\sC} (A_f)) 
	&= \boL\varepsilon_\sC (A_{\settl{\bullet}{A_f}}) \\
	&= A_{\varepsilon_\sC (\settl{\bullet}{A_f}) } \\
	&= A_{\varepsilon_z \name{f} \wr z}, \quad z:A_{\bar f}, \\
	&= A_{\varepsilon_z f\cdot z} \\
	&= A_f.
\end{align*}
Next, let $\phi$ be a term of $\boL\sC$, that is, a polynomial over $\sC$ in variables $x_1, \dots, x_n$, say. Then 
\begin{align*}
\boL\varepsilon_\sC (\eta_{\boL\sC} (\phi) ) 
	&= \varepsilon_{\sC(x_1, \dots, x_n)} (\settl{u:\trut}{\phi})  & \text{by Lemma \ref{l.indeterminatevariable}} \\
	&= \varepsilon_{u} \phi 		\\ 
	&= \phi.
\end{align*}
Next, if $\settl{x}{s}$ is an element of $\boC\Lambda$, then we must verify: %
$$\varepsilon_{\boC\Lambda} ( \boC\eta_{\Lambda} \settl{x}{s} ) = \settl{x}{s}.$$
The first case we check is that where $s$ is a bulletin in a variable not equal to that appearing in the symbol. %
Then 
$\settl{x}{s} = \settl{x}{y} = \ty(x) \vdash \ty(y)$ 
for some variable $y$, with $\ty(y) = \ty(x)$. We have
\begin{align*}
\varepsilon_{\boC\Lambda} (\boC \eta_\Lambda \settl{x}{s} ) 
	&= \varepsilon_{\boC\Lambda} ( \settl{\ksi}{\ksi'} ) , \quad \text{where $\ksi:\settl{\bullet}{\ty(s)}$, $\ksi':\settl{\bullet}{\ty(x)}$} \\
	&= \hat{\ksi} \vdash \hat{\ksi'} \\
	&= \settl{x}{y} \\
	&= \settl{x}{s}.
\end{align*}
Next, we check when %
$s = k$ is a constant term of type $B$ in $\Lambda$, and $x$ is a variable of type $A$ in $\Lambda$. Then 
\begin{align*}
\varepsilon_{\boC\Lambda} (\boC\eta_\Lambda \settl{x}{k} )
	&= \varepsilon_{\boC\Lambda} \settl{\eta_\Lambda x}{\eta_\Lambda k} \\
	&= \varepsilon_{\boC\Lambda} \settl{\ksi}{\settl{u}{k}}, 
		\text{ where $\ksi$ has type $\eta_\Lambda \ty(x)  = \settl{\bullet}{A} = \settl{v:\trut}{A \wr v}$, and $u:\trut_\Lambda$} \\
	&= \varepsilon_\ksi \settl{u}{k} \\
	&= \settl{u}{k} \cdot \arter_{\hat \ksi} \\
	&= \settl{u}{k} \cdot \settl{w:A}{*} \\
	&= \settl{w:A}{k[x/*]} \\
	&= \settl{w}{k} \\
	&= \settl{x}{k}.
\end{align*}
Next, if $s = x$ is a variable of type $A$ and is the same variable as that appearing in the symbol, then
\begin{align*}
\varepsilon_{\boC\Lambda} (\boC\eta_\Lambda \settl{x}{x} )
	&= \varepsilon_{\boC\Lambda} \settl{\eta_\Lambda x}{\eta_\Lambda x} \\
	&= \varepsilon_{\boC\Lambda} \settl{\ksi}{\ksi}, \quad \text{ where $\ksi$ has type $\settl{\bullet}{A} = A$},\\
	&= \varepsilon_\ksi \ksi \\
	&= \settl{\ksi}{\ksi} \\
	&= \settl{x}{x}.
\end{align*}
The other cases are proved similarly. Hence the triangle laws hold, and the theorem is proved. 
\Prf

\section{Conclusion and Future Work}\label{s.conc}

We have shown that cartesian closed structure can be modified to include mappings on the set of objects that recover the base and the power of an exponential. %
We have indicated that the mathematics of cartesian closed categories is not affected by this addition, and moreover, by making this modification, we widen further the class of admissible functors (for some purposes relevant to categorical logic and type theory) %
to include arbitrary cartesian functors. We have also shown that this calculus extends beyond categories, to the generalized categories of section \ref{s.gencat}. 
We have also presented a lambda calculus which permits the extension of the Curry-Howard-Lambek correspondence to the general case. % 
Our work suggests that polynomials over categories and terms over types are in fact essentially the same thing. This can also be seen also in the ordinary categorical case, but in the generalized setting, the observation is made unavoidable. % 
The fundamental insight of the Curry-Howard correspondence is thus that the cartesian closed structure on a cartesian closed category can be expressed almost entirely in terms of properties of objects in the space of polynomials. This seems to be the mathematical content of the theorem. 

Because of the rich variety of subject matter in categorical logic and related subjects, there are a number of directions in which this work can be continued. %
For example, the work of Moggi on computational effects \cite{MoI1} has had an influence on much subsequent work (see for example Wadler\cite{WaR2}, Mulry, \cite{MuY1,MuY2}, Kobayashi \cite{Kobayashi1}, Semmelroth and Sabry \cite{SemmelrothSabry1}). In \cite{MoI1}, an extension of the lambda calculus is introduced and it is shown that it is possible to provide categorical semantics for computational effects by making use of monads. %
In fact, two constructions are presented. The first relates a cartesian closed category equipped with a monad to a monadic equational theory (one in which contexts consist of a unique typed variable) extended by a computational effect (he calls this the {\em simple metalanguage}), and the second relates a {\em strong} monad to a general equational theory (what Moggi calls an {\em algebraic} equational theory, this one called the {\em metalanguage}), i.e., one in which contexts may be arbitrary finite lists of typed variables. %

Let $\sC$ be a category with a monad $T = (T, \eta, \mu)$. Then $T$ is a {\em strong monad} if it is equipped with a natural transformation $t$ from the functor $(-) \times T(-):  \sC \times \sC \to \sC \times \sC$ to the functor $T(- \times -) :  \sC \times \sC \to \sC \times \sC$ (where $\times$ denotes both the product in Cat and the product in $\sC$). This $t$, called a {\em strength}, must additionally satisfy the identities:
$$(T \of \pi_{1,A}) \vertof t_{1, A} = \pi'_{1, TA},$$
$$(T \of \alpha_{A,B,C} ) \vertof t_{A \times B, C} = t_{A, B \times C} \vertof (1_A \times t_{B, C}) \vertof \alpha_{A, B, TC}$$
$$t_{A,B} \vertof (1_A \times \eta_B) = \eta_{A \times B}$$
$$t_{A,B} \vertof (1_A \times \mu_B) = \mu_{A \times B} \vertof (T \of t_{A,B}) \vertof t_{A, TB}$$
where notation is the same as in the preceding sections, except $\times$ denotes the product in $\sC$. 
In \cite{ScMd}, the fundamental parts of the theory of monads are extended to the setting of generalized categories in two ways, one via a generalized triple, and the other via a generalized Kleisli construction. %
Questions remain about how the present work is connected to Moggi's, since the Kleisli category in the generalized setting \cite{ScMd} is a more subtle construction than in the one-categorical setting. 

It is also possible to extend our work in this paper to the setting of topos theory. We may define:

\dfn
An {\em ideal elementary topos} 
is an ideal cartesian closed category with 
\begin{enumerate}
	\item all finite limits and colimits,
	\item a subobject classifier.
\end{enumerate}
\Dfn

\noindent An investigation into topos theory in the generalized setting (including several sheaf theoretical constructions) has been made. % 

\bibliographystyle{abbrv}
\bibliography{mathmain}

\vspace*{10pt}
\noindent {\footnotesize LUCIUS T. SCHOENBAUM \\
\hspace*{18pt}\address{DEPARTMENT OF MATHEMATICS\\
\hspace*{18pt}LOUISIANA STATE UNIVERSITY\\
\hspace*{18pt}BATON ROUGE, LA 70803\\
{\it E-mail}: lschoe2@lsu.edu\\
{\it URL}: \url{http://www.math.lsu.edu/~lschoe2}}
\clearpage

\end{document}